\documentclass[publish,JACIII,paper]{jaciiiarticle}
\usepackage[dvips]{graphicx} 
\usepackage{epsfig}  
\usepackage{amsmath}
\usepackage{subfigure}
\SetVolumeNumber{ } 
\SetArticleID{ }

\begin{document}

\pagestyle{jaciiistyle}

\title{A Method for Accelerating the HITS Algorithm}
\author{Andri Mirzal and Masashi Furukawa}
\address{Graduate School of Information Science and Technology,\\
         Hokkaido University, Kita 14 Nishi 9, Kita-Ku, Sapporo 060-0814, Japan}
\markboth{Mirzal, A., Furukawa, M.}{A Method for Accelerating the HITS Algorithm}

\maketitle

\begin{abstract}% Do not delete this percent symbol
\noindent Abstract. We present a new method to accelerate the HITS algorithm by exploiting hyperlink structure of the web graph. The proposed algorithm extends the idea of authority and hub scores from HITS by introducing two diagonal matrices which contain constants that act as weights to make authority pages more authoritative and hub pages more hubby. This method works because in the web graph good authorities are pointed to by good hubs and good hubs point to good authorities. Consequently, these pages will collect their scores faster under the proposed algorithm than under the standard HITS\@. We show that the authority and hub vectors of the proposed algorithm exist but are not necessarily be unique, and then give a treatment to ensure the uniqueness property of the vectors. The experimental results show that the proposed algorithm can improve HITS computations, especially for back button datasets.
\end{abstract}


\section{Introduction} \label{introduction}
HITS (\emph{Hypertext Induced Topic Search}) is a ranking algorithm introduced by Jon Kleinberg in 1998 \cite{Kleinberg} that utilizes web graph's hyperlink structure to create two metrics associated with every page. The first metric, authority, determines page's popularity, and the second metric, hub, is used to find portal pages, pages that link to popular (thus useful) pages. Because it is easy to create many hyperlinks on a page to boost its hub score (thus can increase authority scores of other pages that are pointed to by it), HITS is susceptible to \emph{link spamming} problem.

HITS is usually being compared to PageRank \cite{Page}, a popular ranking algorithm used by Google that also uses the hyperlink structure to create a popularity measure. Both algorithms were breakthrough achievements at that time because unlike previous methods which usually use page's contents, these algorithms take different approach by utilizing the hyperlink structure to measure page's values. However, there are two main differences that should be enlisted here. \emph{First}, while HITS produces two metrics, PageRank only produces one, the popularity measure. \emph{Second}, unlike PageRank, HITS is \emph{query-dependent}; for every incoming query the algorithm first finds relevant pages (usually by matching terms in the query with the contents), builds neighborhood graph, and then calculates authority and hub scores for every page in the graph. The neighborhood graph is built by not only taking the relevant pages as the vertices, but also other pages that either point to or are being pointed to by the relevant pages. This expansion step allows semantic association to be made and usually solves synonym problem \cite{Kleinberg,Langville}. Unfortunately it also creates famous problem associated with HITS; \emph{topic drift}, authoritative yet irrelevant pages are likely to be also included \cite{Kleinberg,Langville}.

The link spamming problem can be alleviated by giving only fractional weights to edges from mutually reinforcement hosts \cite{Bharat}. The topic drift can be mitigated by computing the relevancy between the query and the pages in the neighborhood graph \cite{Bharat}; the more similar the pages to the query, the more weights they have. So the influence of less relevant pages can be reduced. 

The query-dependence is considered to be the most problematic aspect of HITS because authority and hub vectors have to be calculated online and real time for every incoming query, thus consuming too much computational, memory, and network resources. This problem can be handled by modifying HITS to be query-independent; taking the entire web graph as the neighborhood graph and calculating a global authority and a global hub vector \cite{Langville}. However, some crucial problems faced by PageRank in the early development like storage issues, memory management, tasks division, parallelization strategies, and computational methods must be addressed before this task becomes possible. Fortunately, mathematically query-independent HITS (QI-HITS) resembles PageRank \cite{note1}, so it can be expected that infrastructures and methods built for PageRank can be adopted to QI-HITS.

The challenge of accelerating QI-HITS is not a trivial problem. There are several good reasons to put some efforts on it. \emph{First}, QI-HITS has some nice properties: (1) like PageRank, it can be calculated offline so the system doesn't have to deal with every incoming query and some resources can be saved. (2) Unlike PageRank, it gives two measures; authority scores for finding popular pages and hub scores for finding portal pages. And (3) it solves completely the topic drift problem and slightly reduces the link spamming problem. \emph{Second}, as the web graph is growing rapidly the needs for faster methods are inevitable. For example to keep the freshness of web indices, to save the resources, and to build personalized and topic-sensitive schemes as in the PageRank case \cite{Haveliwala}, among others. Yet to the best of our knowledge, the researches on accelerating HITS are hardly known, partly because HITS is originally query-dependent; the sizes of the neighborhood graph (generally about 1000-5000 pages in 1998 \cite{Kleinberg}) are much smaller than the size of the web graph. Thus, faster and more memory-intensive methods based on matrix inversion or decomposition, especially for sparse symmetric systems can be used \cite{Parlett}. And for QI-HITS (where the problem's scale is as enormous as the PageRank's), some techniques to accelerate PageRank computations like Extrapolation methods \cite{Kamvar}, BlockRank algorithm \cite{Kamvar2}, Gauss-Seidel method \cite{Arasu}, and Reordering methods \cite{Lee,Langville2} can be adopted because both algorithms involving solving dense vector $\times$ sparse matrix operations \cite{note1}.

In this paper, we propose a different approach to accelerate the HITS algorithm. Unlike other methods where the acceleration is gained by using techniques borrowed from linear algebra (Extrapolation and Gauss-Seidel), exploiting adjacency matrix sparseness (Reordering), or utilizing nested block structure of the web graph (BlockRank), this method makes use of the HITS definition itself. So, it can be called \emph{definition-based acceleration method}. 

The proposed algorithm introduces two diagonal matrices, $\mathbf{Ca}$ and $\mathbf{Ch}$, which contain constants associated with authority and hub scores respectively for every page that act as weights to make authority pages more authoritative and hub pages more hubby. Because in the web graph authoritative pages tend to be pointed to by hubby pages, and hubby pages tend to point to authoritative pages, the constants will make these pages collect their scores faster and the stationary distributions (authority and hub vectors) can be reached with less iteration steps. Because this approach makes use of the authority and hub pages, it can be expected that its performance will be better in a graph with considerable proportions of authority and hub pages than in a graph with uniform degree distributions. As shown in previous works \cite{Kleinberg2,Albert,Broder}, the web graph does have power law distributions for both indegrees and outdegrees, so authority and hub pages exist. 

Note that we concern only the performance gained when the approach is applied in conjunction with the power method because in the scale of web graph (1 trillion unique urls in July 2008 based on Google report) only matrix-free iterative methods like the power method, Jacobi, or Gauss-Seidel are feasible to be implemented. Out of these methods, the power method is preferable because it is the simplest, needs less memory, and is linearly scalable to the problem's size. Further, some promising techniques like Reordering and BlockRank are based on the power method.

\section{Related Works} \label{relatedworks}
Some methods to accelerate the PageRank computations that can also be implemented in QI-HITS with some modifications are discussed in this section.

Haveliwala \cite{Haveliwala2} suggests using induced ordering from the  PageRank vector rather than the residual as the stopping criterion, and shows in 24-million-page Stanford WebBase archive dataset the ordering induced by only 10$\mathrm{^{th}}$ iteration agrees fairly well with the ordering induced by 100$\mathrm{^{th}}$ iteration for query specific case. And in the case of global ordering only 25 iterations are needed.

Arasu et.~al.~\cite{Arasu} propose using Gauss-Seidel method instead of the power method. This method immediately updates the entries of the current iteration vector as they become available, thus it clearly converges faster than the power method. A nice thing about this method is its formulation resembles the power method's, so it can easily be implemented with small modifications to the system.

Kamvar et.~al.~introduce two Extrapolation methods to accelerate the PageRank computations \cite{Kamvar}. The methods assume the PageRank vector can be written as linear combination of the eigenvectors of the Google matrix, a stochastic and primitive version of the adjacency matrix induced from the web graph. Because the vector is the principal eigenvector \cite{note2} of the Google matrix, the PageRank convergence can be speeded up by subtracting some subdominant eigenvectors from the current iteration vector. The first method, Aitken Extrapolation, uses successive intermediate vectors to estimate the second eigenvector, and subtracting it from the current iteration vector. The second method, Quadratic Extrapolation, estimates not only the second but also the third eigenvector, and subtracting them from the current iteration vector. It has been shown that the Quadratic Extrapolation is better than the Aitken Extrapolation not only because this method substracts more error, but also there are cases where the second and the third eigenvectors are the repeated vectors, so significant improvement can only be achieved by subtracting both vectors from the current iteration vector.

In their next work, Kamvar et.~al.~propose a very promising aggregation method called BlockRank that speeds up the computation of PageRank by a factor of two times in realistic scenarios \cite{Kamvar2}. This method works well because the web graph has a nested block structure; most pages within a host intralink to other pages within the host, and only a few are interhost links. The method first calculates local PageRank vector for each host by ignoring the interhost links. And a global PageRank vector of hostgraph, a graph created by taking the hosts as the vertices and the interhost links as the edges, is calculated. The global PageRank vector is then used to weight the corresponding local PageRank vectors. The result is taken as a starting vector for the standard PageRank algorithm. Because of this locality approach, BlockRank favors parallelization scheme, thus is very suitable to be implemented in the real condition.

Reordering \cite{Lee,Langville2} is another very promising method to speed up the PageRank calculations, both in the costs per iteration and the number of iterations. The improvement achieved by this method cannot be worse than the original algorithm, and in some datasets, the speedup can reach a factor of 5 times \cite{Langville2}. This method works by exploiting dangling pages, pages with no outlink that usually make up over 80\% of the webpages. In the adjacency matrix representation the dangling pages are the zero rows, so they look alike and can be lumped together into a teleportation state. Consequently, the problem turns into solving PageRank for nondangling pages \cite{Lee}. More recent work by Langville and Meyer \cite{Langville2} provides linear algebra approaches to this problem. They suggest reordering adjacency matrix so that the rows corresponding to the dangling pages are placed at the bottom of the matrix, then the PageRank vector is computed only for nondangling portion. The scores of the dangling pages are recovered by using the vector of nondangling pages and forward substitution.

\section{Proposed Algorithm} \label{proposedalgorithm}
\subsection{Formulation} \label{formulation}
\begin{table}[t]
 \renewcommand{\arraystretch}{1.3}
  \begin{center}
    \caption{Similarity measures between vectors.}
    \centering
    \footnotesize{
    \begin{tabular}{|l|cc|cc|}
    \hline
    Data & \multicolumn{2}{c|}{Auth.~- Indeg.} & \multicolumn{2}{c|}{Hub - Outdeg.} \\
    \cline{2-5}
                 & Cosine & Spearman & Cosine & Spearman \\
    \hline
    britannica.com     & 0.9776 & 0.9614 & 0.9558 & 0.9409 \\
    jobs.ac.uk         & 0.9981 & 0.7590 & 0.5326 & 0.9828 \\
    opera.com          & 0.9337 & 0.5377 & 0.5878 & 0.9904 \\
    python.org         & 0.8629 & 0.6212 & 0.1943 & 0.9722 \\
    scholarpedia.org   & 0.9999 & 0.9986 & 0.6339 & 0.9983 \\
    stanford.edu       & 0.8662 & 0.3794 & 0.5968 & 0.9635 \\
    en.wikipedia.org   & 0.9452 & 0.7323 & 0.8306 & 1.0000 \\
    yahoo.com          & 0.5654 & 0.4968 & 0.4158 & 0.9950 \\
    \hline
    Average      & 0.8936 & 0.6858 & 0.5934 & 0.9804 \\
    \hline
    \end{tabular}}
    \label{table1}
  \end{center}
\end{table}

We will first review the definition and mathematical model of HITS before deriving the proposed algorithm formulation. HITS is defined with the following statement: \emph{authority score of a page is the sum of hub scores of others that point to it, and hub score of a page is the sum of authority scores of others that are pointed to by it} \cite{Kleinberg}. This is a circular statement, the authority scores depend on the hub scores and vice versa. To solve it, every page must be given initial scores, and final scores are computed by successively repeating the summing processes with normalization until a predefined criterion is satisfied. The following equation gives the formulation of HITS:
\begin{equation}
a^{(k+1)}_{i} = \sum_{j\in \mathcal{B}_i} h^{(k)}_{j},\enspace\text{and} \enspace h^{(k+1)}_{i} = \sum_{j\in \mathcal{F}_i} a^{(k+1)}_{j}
\label{eq1}
\end{equation}
for $k = 0,1,\ldots,K-1$, where $K$ denotes the final iteration where the predefined criterion is satisfied, $a_{i}^{(k)}$ and $h_{i}^{(k)}$ denote the authority and the hub score of page $i$ at iteration $k$, $\mathcal{B}_i$ is the set of pages that point to $i$, and $\mathcal{F}_i$ is the set of pages that are pointed to by $i$.

As shown in eq.~\ref{eq1}, HITS simply calculates the authority (hub) score of a page by adding hub (authority) scores of other pages that point to (are pointed to by) it. This original formulation misses an underlying important aspect of the preferential attachment in the web graph; \emph{the portal pages (pages with many outlinks) tend to point to the popular pages (pages with many inlinks), and the popular pages tend to get many new inlinks}. This preferential attachment is the main reason behind the skewed distributions of indegrees and oudegrees as reported in many experiments \cite{Kleinberg2,Albert,Broder}, and because the authority (hub) score of a page is correlated to its indegree (outdegree) \cite{Ding,Ding2}, it can be expected that the stationary distributions of the authority and the hub scores are also skewed. We confirm this fact empirically by calculating the similarities between authority vs.~indegree vectors and hub vs.~outlink vectors. Table \ref{table1} shows the results, while cosine criterion measures the distance between two vectors, Spearman correlation measures the similarity between orderings induced from the vectors (see section \ref{datasets} for details about the datasets).
\begin{figure}[t]
\centering
\includegraphics[width=7cm]{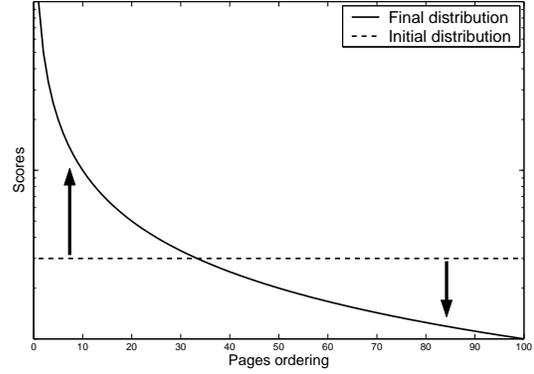}
\caption{The distances between initial and final distributions.}
\label{fig1}
\end{figure}

Usually, a uniform distribution is used as the starting vector. Thus, the distances between initial and final scores are not uniform. For some very authoritative and hubby pages, it takes more iteration steps to reach the final scores. This is also true for pages that have very low final authority or hub scores. Fig.~\ref{fig1} describes such condition; the distances between initial and final scores of pages that ordered in the top and bottom are greater than pages in the middle positions.

The proposed algorithm is formulated to deal with this problem. It measures the distances between initial and final scores, and sets the convergence velocities proportional to the distances. As stated earlier the final authority and hub scores can be roughly approximated by using indegree and outdegree distributions, so it will be utilized to create constants that determine the convergence velocities.

Let $ca_i$ and $ch_i$ be the constants associated with the authority and hub score of page $i$, and make some observations before writing down the formulations. Clearly, $ca_i$ must be bigger than $ch_i$ if $i$ has many inlinks than outlinks, $ch_i$ must be bigger than $ca_i$ if $i$ has many outlinks than inlinks, and $ca_i$ must be equal to $ch_i$ if $i$ has the same number of inlinks and outlinks. Also, the addition of a new link to pages with small number of links should have greater impact than highly connected pages. By using these observations, we define the constants as:
\begin{eqnarray}
ca_{i} &=& \frac{\text{indeg}_{i}}{\text{deg}_{i}}
       |\text{indeg}_{i} - \text{outdeg}_{i}|
        ^{p_{i}}, \enspace \text{and} \enspace\enspace\enspace \\
ch_{i} &=& \frac{\text{outdeg}_{i}}{\text{deg}_{i}}
       |\text{indeg}_{i} - \text{outdeg}_{i}|
        ^{-p_{i}}, \; \text{where} \\
p_{i} &=& \left\{
 \begin{array}{rl}
  1  & \text{if } \text{indeg}_{i} > \text{outdeg}_{i} \\
  -1 & \text{if } \text{indeg}_{i} < \text{outdeg}_{i} \\
  0  & \text{otherwise}
 \end{array} \right. \nonumber
\end{eqnarray}
where indeg$_i$, outdeg$_i$, and deg$_i$ denote the indegree, outdegree, and degree of $i$ respectively, and $|\ast|$ denotes the absolute value of $\ast$. And the proposed algorithm is defined with the following equation:
\begin{equation}
a^{(k+1)}_{i} = \sum_{j\in \mathcal{B}_i} h^{(k)}_{j}ch_{j}, \enspace \text{and} \enspace
h^{(k+1)}_{i} = \sum_{j\in \mathcal{F}_i} a^{(k+1)}_{j}ca_{j}
\label{eq4}
\end{equation}

As shown above, $h_{j}$ and $a_{j}$ are weighted with $ch_{j}$ and $ca_{j}$ respectively. Because $ca_{j}$ and $ch_{j}$ are proportional to indeg$_i$ and outdeg$_i$ which in turn are likely to be proportional to the distances between final and initial authority and hub vectors, these constants tend to make the portal and popular pages and also the pages located in the bottom positions collect their scores faster as the iteration steps progress. Thus, it can be expected that the proposed algorithm will converge faster than HITS\@.

The proposed algorithm will be represented in matrix notation for some reasons: (1) to simplify the formulation, (2) to allow graph properties being seen in linear algebra perspective, (3) to compare its formulation to the HITS's and PageRank's, (4) to allow other acceleration methods stated previously be applied with ease, and (5) to analyze the convergence property (see section \ref{convergenceAnalysis}). Let $\mathbf{L}$ be the adjacency matrix of the web graph where $\mathbf{L}_{ij}$ is 1 if there is a link from $i$ to $j$, and 0 otherwise, $\mathbf{Ca} = \text{diag}(ca_1, ca_2, \ldots, ca_N)$, $\mathbf{Ch} = \text{diag}(ch_1, ch_2, \ldots, ch_N)$, and $N$ is the number of pages in the web graph. Thus, the proposed algorithm can be rewritten as:
\begin{equation}
\mathbf{a}^{(k+1)T} = \mathbf{h}^{(k)T}\mathbf{Ch}\,\mathbf{L},
\enspace \text{and} \enspace
\mathbf{h}^{(k+1)T} = \mathbf{a}^{(k+1)T}\mathbf{Ca}\,\mathbf{L}^T
\label{eq5}
\end{equation}
where $\mathbf{a}^{T}$ is 1$\times N$ authority vector and $\mathbf{h}^{T}$ is 1$\times N$  hub vector.

Algorithm 1, 2, and 3 are used to calculate QI-HITS, the proposed algorithm, and PageRank respectively, where $\|\ast\|_1$ denotes 1-norm of $\ast$, $\epsilon$ denotes the desired residual level, $\mathbf{p}^T$ denotes the 1$\times N$ PageRank vector, $\mathbf{Do}=\text{diag}(\text{outdeg}_1,\text{outdeg}_2,\ldots,\text{outdeg}_N)$ denotes the diagonal outdegree matrix, $\mathbf{d}$ denotes $N\times$1 dangling vector where its $n^{\mathrm{th}}$ ($n=1,2,\ldots,N$) entry is 1 if $n$ is a dangling page and 0 otherwise, and $0<\alpha<1$ denotes a scalar that controls the proportion of time a random surfer follows the hyperlinks as opposed to teleporting. Algorithm 3 is adopted from works by Langville. Detail discussions can be found in \cite{Langville,Langville3}.
\begin{table}[t]
 \renewcommand{\arraystretch}{1.3}
 \centering
  \small{
  \subtable{
   \centering
    \begin{tabular}{|l|}
    \hline
    \textbf{Algorithm 1}: QI-HITS \\
    \hline
    Initialize $\mathbf{h}^{(k=0)T}=\mathbf{e}^T/N$ \\
    While $\delta > \epsilon$, do: \\
    \quad $\mathbf{a}^{(k+1)T} = \mathbf{h}^{(k)T}\mathbf{L}$ \\
    \quad $\mathbf{h}^{(k+1)T} = \mathbf{a}^{(k+1)T}\mathbf{L}^T$ \\
    \quad Normalize $\mathbf{h}^{(k+1)T}$ \\
    \quad $\delta = \|\mathbf{h}^{(k+1)T} - \mathbf{h}^{(k)T}\|_1$ \\
    \quad $k = k + 1$ \\
    Normalize $\mathbf{a}^T$ \\
    Return $\mathbf{a}^T$ and $\mathbf{h}^T$ \\
    \hline
   \end{tabular}
  }

  \subtable{
   \centering
    \begin{tabular}{|l|}
    \hline
    \textbf{Algorithm 2}: Prop.~Alg. \\
    \hline
    Initialize $\mathbf{h}^{(k=0)T}=\mathbf{e}^T/N$ \\
    Calculate $\mathbf{Ca}$ and $\mathbf{Ch}$ \\
    While $\delta > \epsilon$, do: \\
    \quad $\mathbf{a}^{(k+1)T} = \mathbf{h}^{(k)T}\mathbf{Ch}\,\mathbf{L}$ \\
    \quad $\mathbf{h}^{(k+1)T} = \mathbf{a}^{(k+1)T}\mathbf{Ca}\,\mathbf{L}^T$ \\
    \quad Normalize $\mathbf{h}^{(k+1)T}$ \\
    \quad $\delta = \|\mathbf{h}^{(k+1)T} - \mathbf{h}^{(k)T}\|_1$ \\
    \quad $k = k + 1$ \\
    Normalize $\mathbf{a}^T$ \\
    Return $\mathbf{a}^T$ and $\mathbf{h}^T$ \\
    \hline
    \end{tabular}
  }
  \\
  \subtable{
   \centering
    \begin{tabular}{|l|}
    \hline
    \textbf{Algorithm 3}: PageRank \\
    \hline
    Initialize $\mathbf{p}^{(k=0)T}=\mathbf{e}^T/N$ \\
    While $\delta > \epsilon$, do: \\
    \quad $\mathbf{p}^{(k+1)T} = \alpha\mathbf{p}^{(k)T}\mathbf{Do}^{-1}\mathbf{L}+(\alpha\mathbf{p}^{(k)T}\mathbf{d}+1-\alpha)\mathbf{e}^T/N$ \\
    \quad $\delta = \|\mathbf{p}^{(k+1)T} - \mathbf{p}^{(k)T}\|_1$ \\
    \quad $k = k + 1$ \\
    Return $\mathbf{p}^T$ \\
    \hline
    \end{tabular}
  }
 }
\end{table}

\subsection{Operational Costs and Memory Requirements} \label{costs}
In QI-HITS, there are two dense vector $\times$ sparse matrix operations for each iteration step. Because $\mathbf{L}$ contains only either 1 or 0, the cost of each multiplication is \textit{nnz}($\mathbf{L}$) additions, where \textit{nnz}($\mathbf{L}$) denotes the number of nonzero of $\mathbf{L}$. And the normalization step needs $N$ multiplications. Thus, QI-HITS needs $N$ multiplications and 2\textit{nnz}($\mathbf{L}$) additions per iteration. 

In the proposed algorithm, $\mathbf{a}$ and $\mathbf{h}$ are multiplied by \textbf{Ca} and \textbf{Ch} respectively, so there are additional 2$N$ multiplications. Thus, the proposed algorithm needs 3$N$ multiplications and 2\textit{nnz}($\mathbf{L}$) additions per iteration. 

In PageRank, $\mathbf{p}$ is multiplied by $\mathbf{Do}^{-1}$ which needs $N$ multiplications. Then the result is multiplied by $\mathbf{L}$, which needs \textit{nnz}($\mathbf{L}$) additions. Further there are also additional adjustments (stochasticity and primitivity) to ensure the convergence of the result, which need $|$ND$|$ multiplications and ($N$ + $|$ND$|$) additions, where $|$ND$|$ denotes the number of nondangling pages. Because there is no need to do normalization in PageRank, the costs are $N$ + $|$ND$|$ multiplications and (\textit{nnz}($\mathbf{L}$) + $N$ + $|$ND$|$) additions per iteration. Table \ref{table2} summarizes the costs.

The memory requirements for $\mathbf{L}$ is \textit{nnz}($\mathbf{L}$) booleans; $\mathbf{h}^{(k)T}$, $\mathbf{a}^{(k+1)T}$, $\mathbf{h}^{(k+1)T}$, $\mathbf{Ca}$, and $\mathbf{Ch}$ is $N$ doubles each; $\mathbf{Do}^{-1}$ is $N$ integers; and $\mathbf{d}$ is $|\textrm{ND}|$ booleans. Table \ref{table3} summarizes the required memory.
\begin{table}[t]
 \renewcommand{\arraystretch}{1.3}
  \begin{center}
    \caption{Operational costs for original datasets.}
    \centering
    \footnotesize{
    \begin{tabular}{|l|l|l|}
    \hline
    Algorithm & Mult. & Addition \\
    \hline
    QI-HITS   & \textit{N} & 2\textit{nnz}(\textbf{L}) \\
    Prop.~Alg. & 3\textit{N} & 2\textit{nnz}(\textbf{L}) \\
    PageRank & \textit{N} + $|$ND$|$ & \textit{nnz}(\textbf{L}) + \textit{N} + $|$ND$|$ \\
    \hline
    \end{tabular}}
    \label{table2}
  \end{center}
\end{table}
\begin{table}[t]
 \renewcommand{\arraystretch}{1.3}
  \begin{center}
    \caption{Memory requirements for original datasets.}
    \centering
    \footnotesize{
    \begin{tabular}{|l|l|}
    \hline
    Algorithm & Memory \\
    \hline
    QI-HITS    & \textit{nnz}(\textbf{L}) bools and 3\textit{N} doubles \\
    Prop.~Alg. & \textit{nnz}(\textbf{L}) bools and 5\textit{N} doubles \\
    PageRank   & \textit{nnz}(\textbf{L}) + $|\textrm{ND}|$ bools, \textit{N} integers, and 2\textit{N} doubles \\
    \hline
    \end{tabular}}
    \label{table3}
  \end{center}
\end{table}

\subsection{The Back Button Model} \label{backbutton}
The dangling pages can cause computational issues both for PageRank and HITS\@. In the PageRank case, the score of a page is originally defined as the proportion of time the random surfer spends on the page after following the hyperlink structure of the web infinitely \cite{Page}. If the web graph contains the dangling pages, the random surfer will not be able to traverse all pages because it will be trapped in a dangling page if encountering it. Consequently, not only the scores can only be defined for the pages that have been visited, but also due to the finite time of the observation, these scores will not reflect the real values of the pages.

In linear algebra perspective, the dangling pages make the web graph not strongly connected; the adjacency matrix induced from the web graph is reducible. And by Perron theorem for nonnegative matrices \cite{Langville,Farahat}, the dominant eigenvector of a nonnegative reducible matrix exists but is not necessarily positive and unique. Thus, there is no guarantee a unique and positive PageRank vector exists for the original PageRank problem; finding the dominant eigenvector of $(\mathbf{Do}^{-1}\mathbf{L})^{T}$ \cite{note3}. 
%Note that even though the matrix is irreducible (which is very unlikely), by Perron-Frobenius theorem, there is still no guarantee that the dominant eigenvector will be unique due to the possibility of having other eigenvalues with the same moduli to the dominant eigenvalue on the spectral circle \cite{Langville}.

The same condition goes for HITS, both authority matrix $\mathbf{L}^{T}\mathbf{L}$ ($\mathbf{a}^{(k+1)T}=\mathbf{a}^{(k)T}\mathbf{L}^{T}\mathbf{L}$) and hub matrix $\mathbf{L}\,\mathbf{L}^{T}$ ($\mathbf{h}^{(k+1)T}=\mathbf{h}^{(k)T}\mathbf{L}\,\mathbf{L}^{T}$) are nonnegative. Thus by Perron theorem for nonnegative matrices, $\mathbf{a}^{T}$ and $\mathbf{h}^{T}$ exist but there is no guarantee of the uniqueness \cite{note4}.

In addition to the non-uniqueness, there is also problem related to the final distributions as in the PageRank case. The dangling pages receive scores from others that point to them but do not share their scores (because they have no outdegree), so they will always become authoritative and have no hub scores. Consequently, the authority and hub distributions will be more skewed as the number of dangling pages increases, and the average distances between initial distributions (usually uniform distributions) and final stationary distributions in the web graph with many dangling pages are greater than in the web graph without dangling page. Thus, the convergence rates tend to become slower as the number of the dangling pages increases. 
\begin{table}[t]
 \renewcommand{\arraystretch}{1.3}
  \begin{center}
    \caption{Operational costs for back button datasets.}
    \centering
    \footnotesize{
    \begin{tabular}{|l|l|l|}
    \hline
    Algorithm & Mult. & Addition \\
    \hline
    QI-HITS    & \textit{N}  & 2\textit{nnz}(\textbf{L}$^*$) \\
    Prop.~Alg. & 3\textit{N} & 2\textit{nnz}(\textbf{L}$^*$) \\
    PageRank   & \textit{N}  & \textit{nnz}(\textbf{L}$^*$) + \textit{N} \\
    \hline
    \end{tabular}}
    \label{table4}
  \end{center}
\end{table}
\begin{table}[t]
 \renewcommand{\arraystretch}{1.3}
  \begin{center}
    \caption{Memory requirements for back button datasets.}
    \footnotesize{
    \begin{tabular}{|l|l|}
    \hline
    Algorithm & Memory \\
    \hline
    QI-HITS    & \textit{nnz}(\textbf{L}$^*$) bools and 3\textit{N} doubles \\
    Prop.~Alg. & \textit{nnz}(\textbf{L}$^*$) bools and 5\textit{N} doubles \\
    PageRank   & \textit{nnz}(\textbf{L}$^*$) bools, 
                 \textit{N} integers, and 2\textit{N} doubles \\
    \hline
    \end{tabular}}
    \label{table5}
  \end{center}
\end{table}

Our experiments confirm this effect. As shown in Fig.~\ref{fig2}, the PageRank convergence rates are almost the same or better than HITS, which do not agree with previous experiments where the researchers usually remove the dangling pages from the datasets \cite{Kleinberg,Ding,Ding2,Ng,Ng2}. And in the datasets without the dangling page (Fig.~\ref{fig3}), the results agree with the previous experiments; HITS usually converges faster than PageRank.

To deal with the dangling pages, instead of removing them, we prefer to use the back button model \cite{Fagin,Mathieu,Sydow}. This is because \textit{first}, web graph datasets usually have high percentages of the dangling pages, so great portion of useful data will be lost if they are removed. \textit{Second}, some of the dangling pages are the important pages \cite{Eiron}, so removing them can bias the results. And \textit{third}, most users usually go back to the previous page when encountering a dangling page, so this is a natural way in modeling the web graph. Mathematically, the back button model rewrites \textbf{L} into $\mathbf{L}^{\ast}=\mathbf{L}+\mathbf{M}$, where \textbf{M} is $N\times N$ matrix with row \textit{i} is equal to column \textit{i} of \textbf{L} if \textit{i} is a dangling page, and $\mathbf{0}_{1\times N}$ otherwise.

Table \ref{table6} shows the average fractions of authoritative and hubby pages in the original and the back button datasets, where \textit{f\mbox{}i} denotes the average fraction of indegree and \textit{f\mbox{}o} denotes the average fraction of outdegree. The original datasets in average have authoritative pages more than 93\% (\textit{fi} $>$ 0.6) and this percentage is almost unchanged for very authoritative pages (\textit{fi} $>$ 0.9). As shown in Table \ref{table1}, the average percentage of the dangling pages is 92.9\%, so almost all authoritative pages are the dangling pages. Conversely, the number of hubby pages is only about 6\% and is also almost unchange for very hubby pages.
\begin{table*}[t]
 \renewcommand{\arraystretch}{1.3}
  \begin{center}
   \caption{Average fractions of authoritative and hubby pages.}
    \footnotesize{
    \begin{tabular}[t]{|l|rrrr|rrrr|}
     \hline
     Data & \multicolumn{4}{c}{Authoritative page} & 
            \multicolumn{4}{|c|}{Hubby Page} \\ \cline{2-9}
          & \textit{f\mbox{}i} $>$ 0.6 & \textit{f\mbox{}i} $>$ 0.7 & \textit{f\mbox{}i} $>$ 0.8 & 
            \textit{f\mbox{}i} $>$ 0.9 & \textit{f\mbox{}o} $>$ 0.6 & \textit{f\mbox{}o} $>$ 0.7 & 
            \textit{f\mbox{}o} $>$ 0.8 & \textit{f\mbox{}o} $>$ 0.9 \\
     \hline
     Original    & 0.9334 & 0.9320 & 0.9310 & 0.9303 & 
                   0.0661 & 0.0644 & 0.0613 & 0.0556 \\
     Back button & 0.0189 & 0.0041 & 0.0022 & 0.0013 & 
                   0.0503 & 0.0373 & 0.0253 & 0.0102 \\
     \hline
    \end{tabular}}
	\label{table6}
  \end{center}
\end{table*}

When the back button model is applied, the percentages of authoritative pages drop significantly and are comparable to the percentages of hubby pages. Because this model turns the dangling pages into nondangling ones, the remaining pages are the real authority; pages with many indegrees than outdegrees, not pages that have only indegrees.

The costs and memory requirement of QI-HITS, the proposed algorithm, and PageRank for back button datasets are shown in Table \ref{table4} and \ref{table5}.

\subsection{Convergence Analysis} \label{convergenceAnalysis}
To analyze the convergence property of the proposed algorithm, the eq.~\ref{eq5} will be rewritten in one vector $\times$ matrix representation instead of two. Let $\mathbf{X} = \mathbf{Ca}\, \mathbf{L}^{T}\mathbf{Ch}\,\mathbf{L}$, the authority vector of the proposed algorithm can be rewritten into:
\begin{equation}
\mathbf{a}^{(k+1)T} = \mathbf{a}^{(k)T}\mathbf{X}
\label{eq6}
\end{equation}
and after the power method applied to the above equation has converged, the hub vector can be revived by calculating $\mathbf{h}^{T} = \mathbf{a}^{T}\mathbf{Ca}\,\mathbf{L}^{T}$. Consequently, the problem of finding authority and hub vectors can be reduced into only calculating dominant eigenvector of $\mathbf{X}^{T}$ (note that HITS can also be rewritten into this style \cite{note1}).

Because $\mathbf{X}^{T}$ is nonnegative, it always has a nonnegative dominant eigenvalue $\lambda_1$ such that the moduli of all other eigenvalues \{$\lambda_2, \ldots, \lambda_k$\} do not exceed $\lambda_1$, and a dominant eigenvector corresponding to $\lambda_1$ can be chosen so that every entry is nonnegative (see theorem 3.6 in \cite{Farahat}). Thus, the existence of the authority vector of the proposed algorithm is guaranteed. But depending on the initialization, it may be not unique since $\lambda_1$ may be repeated \cite{Langville,Farahat}.

To guarantee the uniqueness of the proposed algorithm, the matrix $\mathbf{X}$ must be modified into a positive matrix. The positive version of $\mathbf{X}$ can be written as: $\mathbf{\hat{X}}=\zeta\,\mathbf{X}+(1/N)(1-\zeta)\,\mathbf{e}\,\mathbf{e}^T$, where $0<\zeta<1$ is a constant that should be set near to $1$ to preserve the hyperlink structure information. And the proposed algorithm can be rewritten as:
\begin{equation}
\mathbf{a}^{(k+1)T} = \mathbf{a}^{(k)T}\mathbf{\hat{X}}
\label{eq7}
\end{equation}
Because $\mathbf{\hat{X}}$ is not stochastic, $\mathbf{a}^{T}$ must be normalized for each iteration step. 

By Perron theorem for positive matrices \cite{Langville,Farahat}, a unique and positive principal eigenvector of $\mathbf{\hat{X}}^{T}$ (the authority vector of the proposed algorithm) is guaranteed to exist. And by ensuring the starting vector not in the range $(\mathbf{\hat{X}}^{T}-\lambda_{1}\mathbf{I})$, the power method applied to eq.~\ref{eq7} is guaranteed to converge to this vector \cite{Langville}. In particular, all positive vectors satisfy the requirement as the starting vector \cite{Chakrabarti}.

Note that in addition to guaranteeing the uniqueness, this modification also tackles the second less obvious problem associated with the proposed algorithm (and also HITS); producing ranking vectors that inappropriately assign zero scores to some pages \cite{Farahat}.

In practice however, as with the HITS case, usually the proposed algorithm can be used without this modification because the scores from the link structure ranking algorithms will be combined with other scores like contents and hypertext scores, making the final scores less sensitive to the ranking vectors. And actually based on our experiments, the HITS and the proposed algorithm do converge to unique nonnegative vectors for all datasets (including the back button datasets) with $\mathbf{e}^{T}/N$ as the starting vector.

\section{Experimental Results} \label{results}
In this section the performance of the proposed algorithm is evaluated by comparing its convergence rates and processing times to reach the same corresponding residual level with the results of QI-HITS and PageRank. The suitability of the proposed algorithm in approximating HITS is confirmed in section \ref{similarity}. And the examples of the top pages returned by the algorithms are given in section \ref{toppages}.

The experiments are conducted by using a notebook with 1.86 GHz Intel processor and 2 GB RAM. The codes are written in python by extensively using database to store lists of adjacency matrix, score vectors, and other related data in harddisk, and open sourced as the part of our work in developing a simple web search engine for research purposes \cite{Mirzal}.

\subsection{The Datasets} \label{datasets}
There are 8 datasets used in the experiments that consist around 10 thousands to 225 thousands pages with average degrees from around 4 to 47. Except Wikipedia \cite{Segaran}, all datasets were crawled by using our crawling system \cite{Mirzal}. All datasets, but Britannica, have a typical web graph’s average degree, around 4 to 15 \cite{Kamvar2,Langville}. However, the percentages of the dangling pages are quite higher here than in a typical dataset (around 70\% to 85\%) due to the high number of downloaded but unexplored pages. Table \ref{table7} summarizes the datasets, where \%DP denotes percentage of the dangling pages, and AD denotes average degree.
\begin{table}[t]
 \renewcommand{\arraystretch}{1.3}
  \begin{center}
   \caption{Datasets summary.}
    \footnotesize{
    \begin{tabular}[t]{|l|l|r|r|r|r|}
     \hline
     Data             & Crawled & Pages & Links & \%DP & AD \\
     \hline
     britannica   & 09/2008    & 21104   & 994554  & 85.0  & 47.1  \\
     jobs         & 12/2008    & 16056   & 187957  & 92.0  & 11.7  \\
     opera        & 12/2008    & 49749   & 437748  & 95.4  & 8.8   \\
     python       & 09/2008    & 57328   & 449529  & 93.5  & 7.8   \\
     scholarpedia & 06/2008    & 74243   & 1077781 & 86.5  & 14.5  \\
     stanford     & 12/2008    & 225441  & 2196441 & 96.7  & 9.7   \\
     wikipedia    & 09/2006 & 10431   & 46152 & 96.1 & 4.4   \\
     yahoo        & 12/2008    & 34054   & 161700  & 98.0  & 4.7   \\
     \hline
     \multicolumn{2}{|c|}{Average} & 61050  & 693983  & 92.9 & 13.6 \\
     \hline
    \end{tabular}}
	\label{table7}
  \end{center}
\end{table}
\begin{table}[t]
 \renewcommand{\arraystretch}{1.3}
  \begin{center}
    \caption{Average similarity measures.}
    \footnotesize{
    \begin{tabular}{|l|rr|rr|}
     \hline
     Data & \multicolumn{2}{c}{Authority} & \multicolumn{2}{|c|}{Hub} \\
     \cline{2-5}
                 & Cosine & Spearman & Cosine & Spearman \\
     \hline
     Original    & 0.859 & 0.810 & 0.976 & 0.999 \\
     Back button & 0.912 & 0.794 & 0.945 & 0.861 \\
     \hline
    \end{tabular}}
   \label{table8}
  \end{center}
\end{table}

\subsection{Convergence Rates} \label{convergence}
Fig.~\ref{britannica}-\ref{yahoo} and \ref{britannicabb}-\ref{yahoobb} show the convergence rates for the original and the back button datasets respectively. The horizontal axis is the number of iteration and the vertical axis is the residual. In the original datasets, the proposed algorithm converges faster than HITS (except for yahoo dataset, where the percentage of the dangling pages is too high), but generally still cannot beat PageRank. In the back button model, where the dangling pages are forced to have outdegree, the proposed algorithm gives very promising results, faster than both HITS and PageRank for all datasets. Further, in the back button model generally HITS converges faster than PageRank, which agrees with the previous works \cite{Kleinberg,Ding,Ding2,Ng,Ng2}.

\subsection{Processing times} \label{times}
\begin{figure*}
 \begin{center}
 
  \subfigure[]{
   \includegraphics[width=0.3\textwidth]{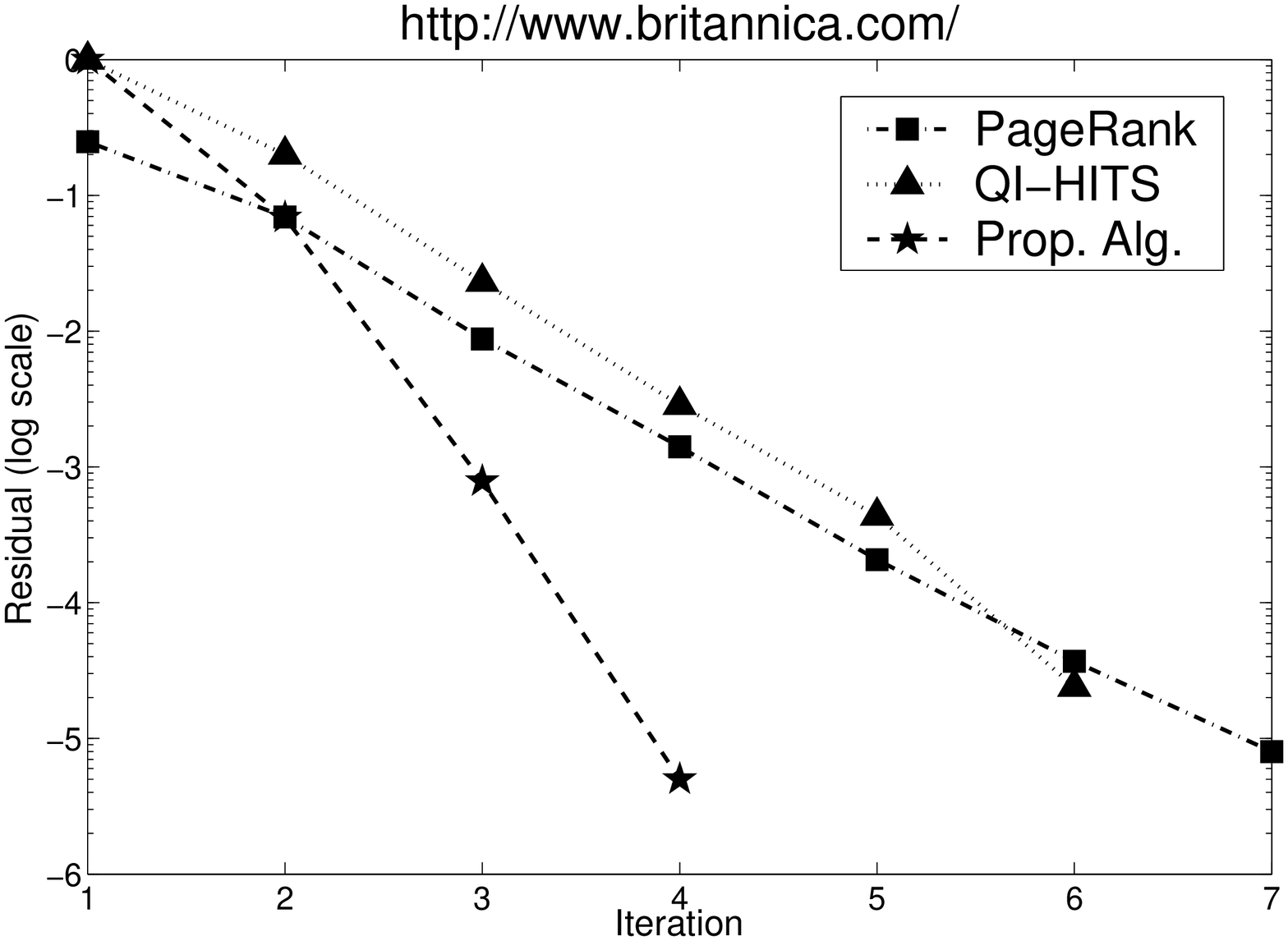}
   \label{britannica}
  }
  \subfigure[]{
   \includegraphics[width=0.3\textwidth]{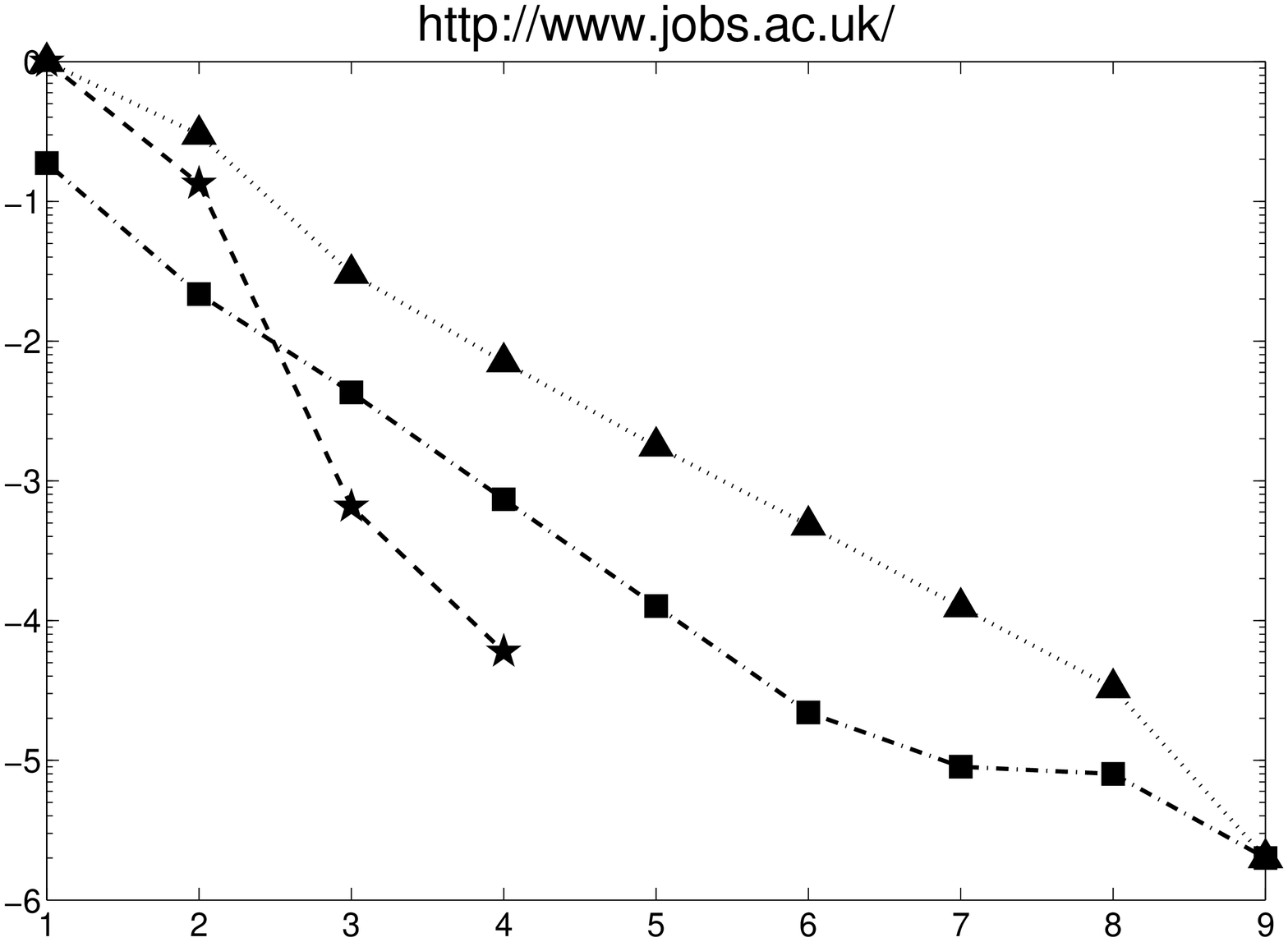}
   \label{jobs}
  }
  \subfigure[]{
   \includegraphics[width=0.3\textwidth]{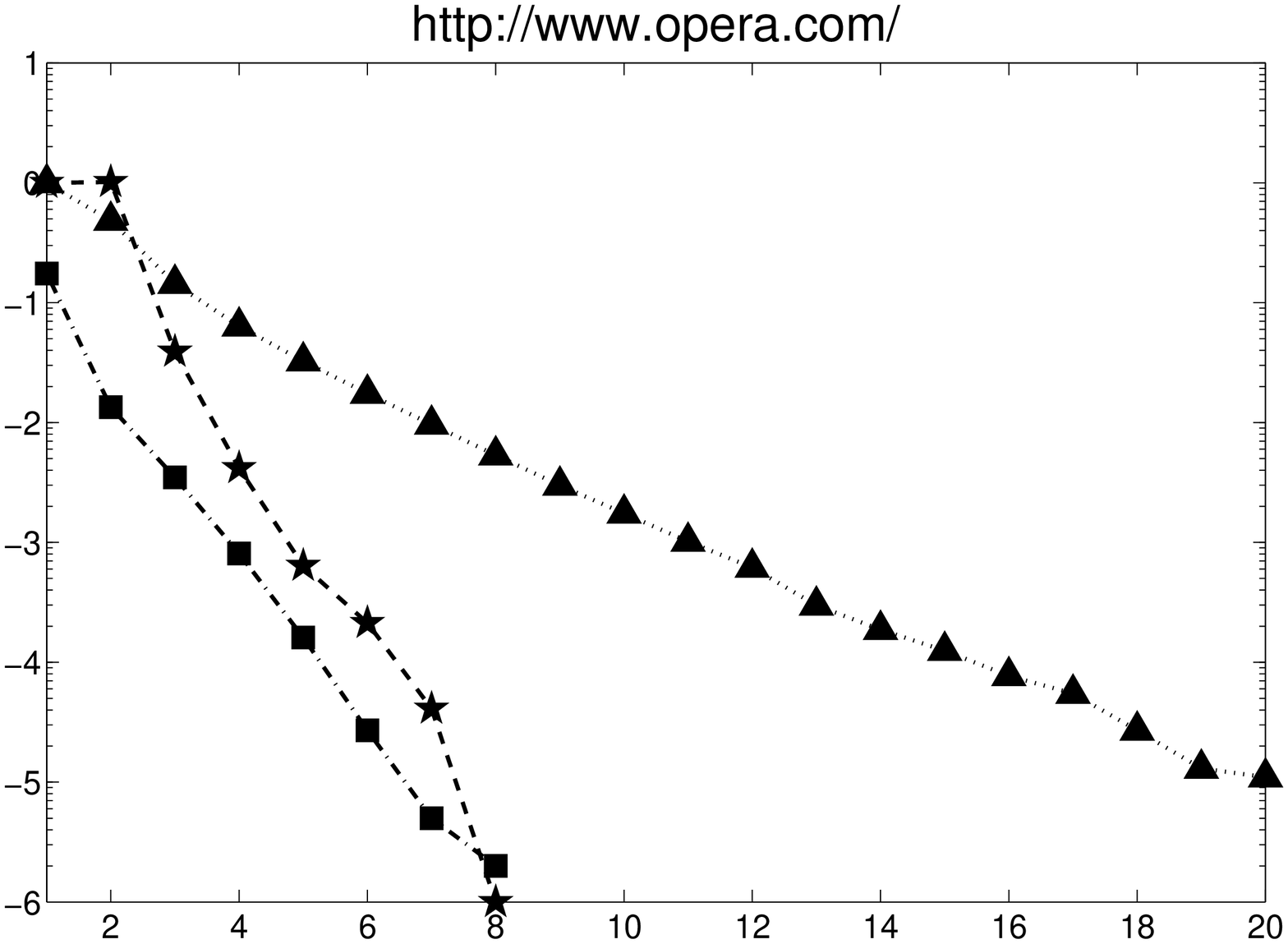}
   \label{opera}
  }
  \\
 
  \subfigure[]{
   \includegraphics[width=0.3\textwidth]{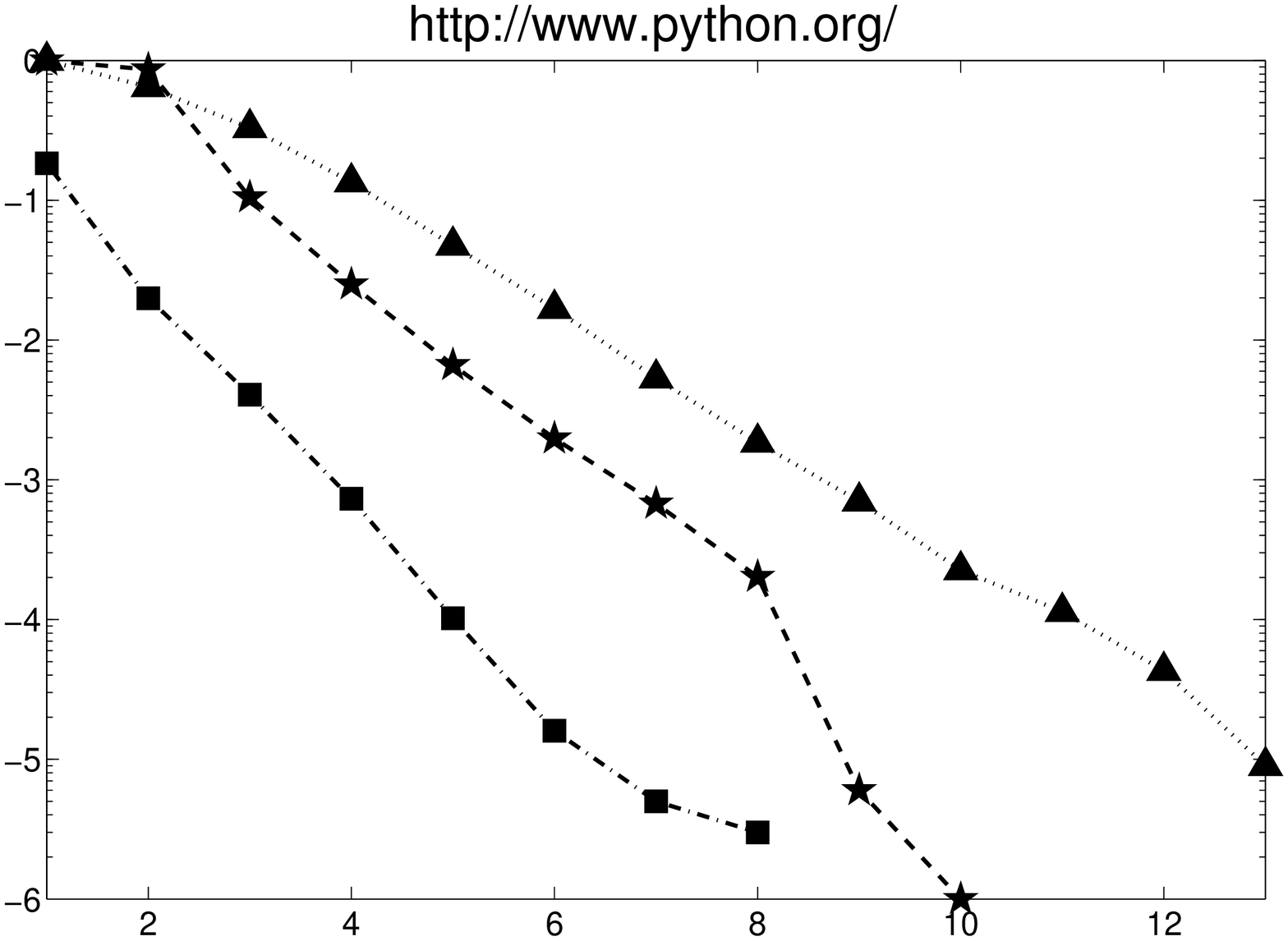}
   \label{python}
  }
  \subfigure[]{
   \includegraphics[width=0.3\textwidth]{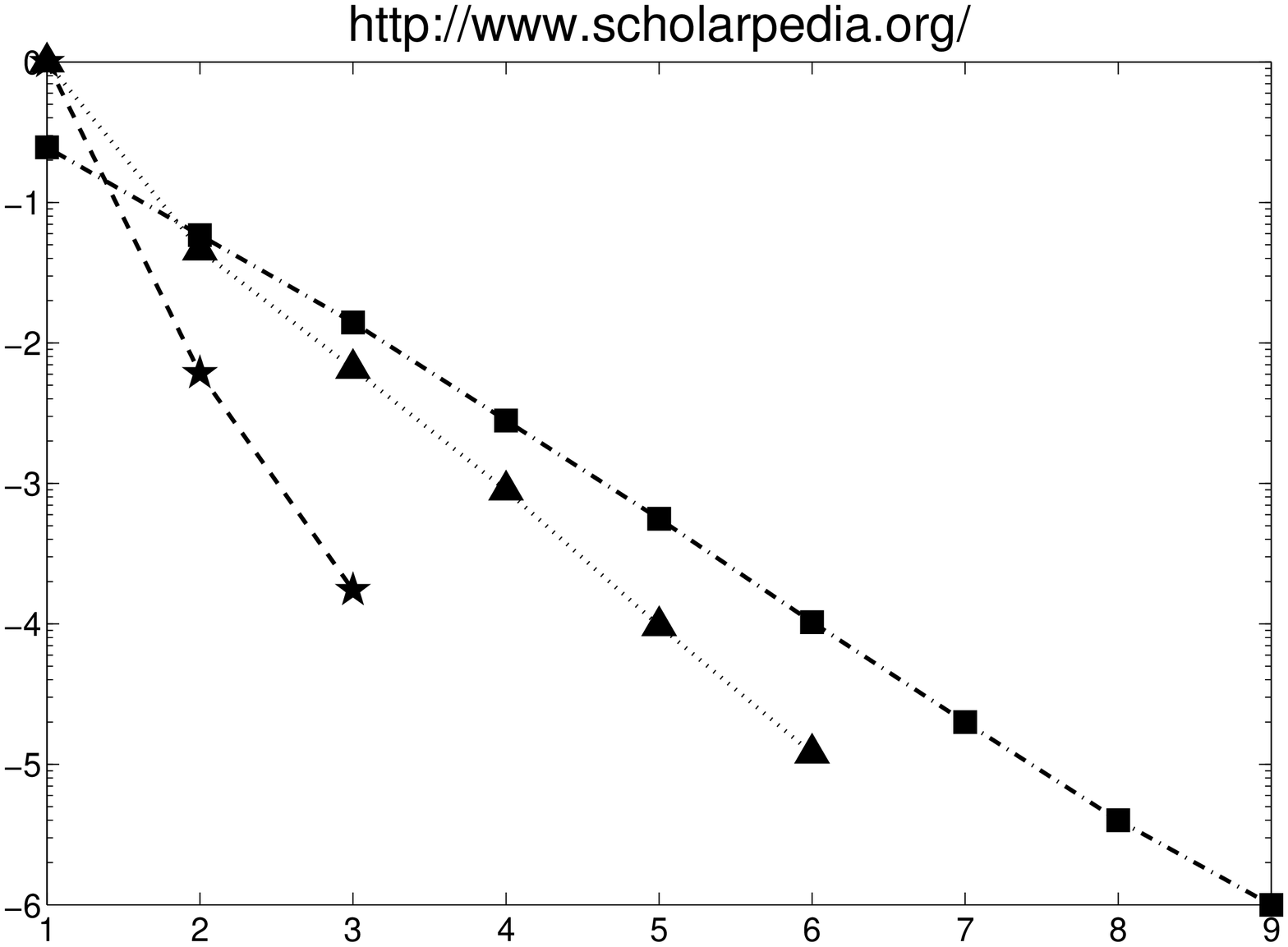}
   \label{scholarpedia}
  }
  \subfigure[]{
   \includegraphics[width=0.3\textwidth]{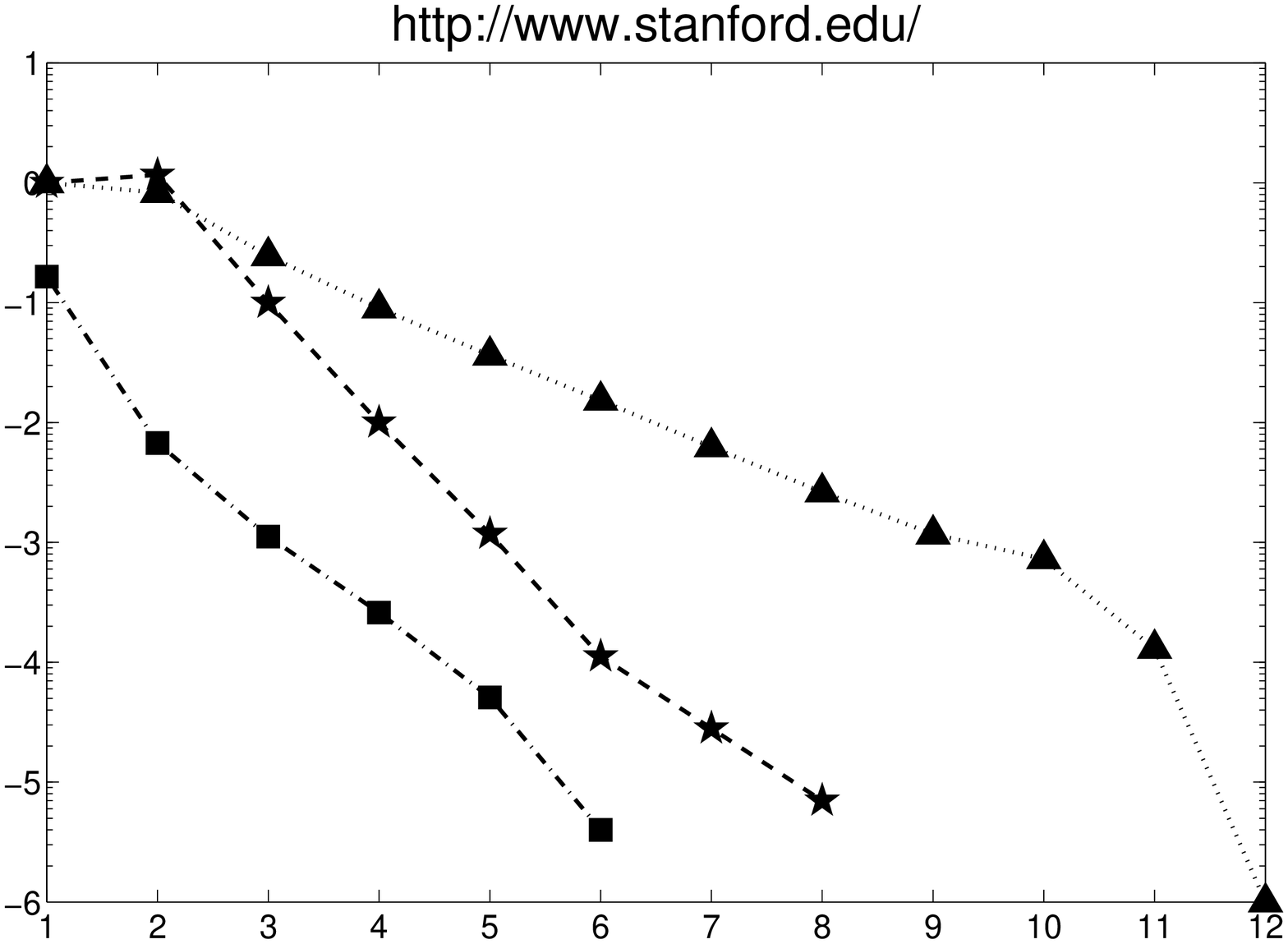}
   \label{stanford}
  }
  \\

  \subfigure[]{
   \includegraphics[width=0.3\textwidth]{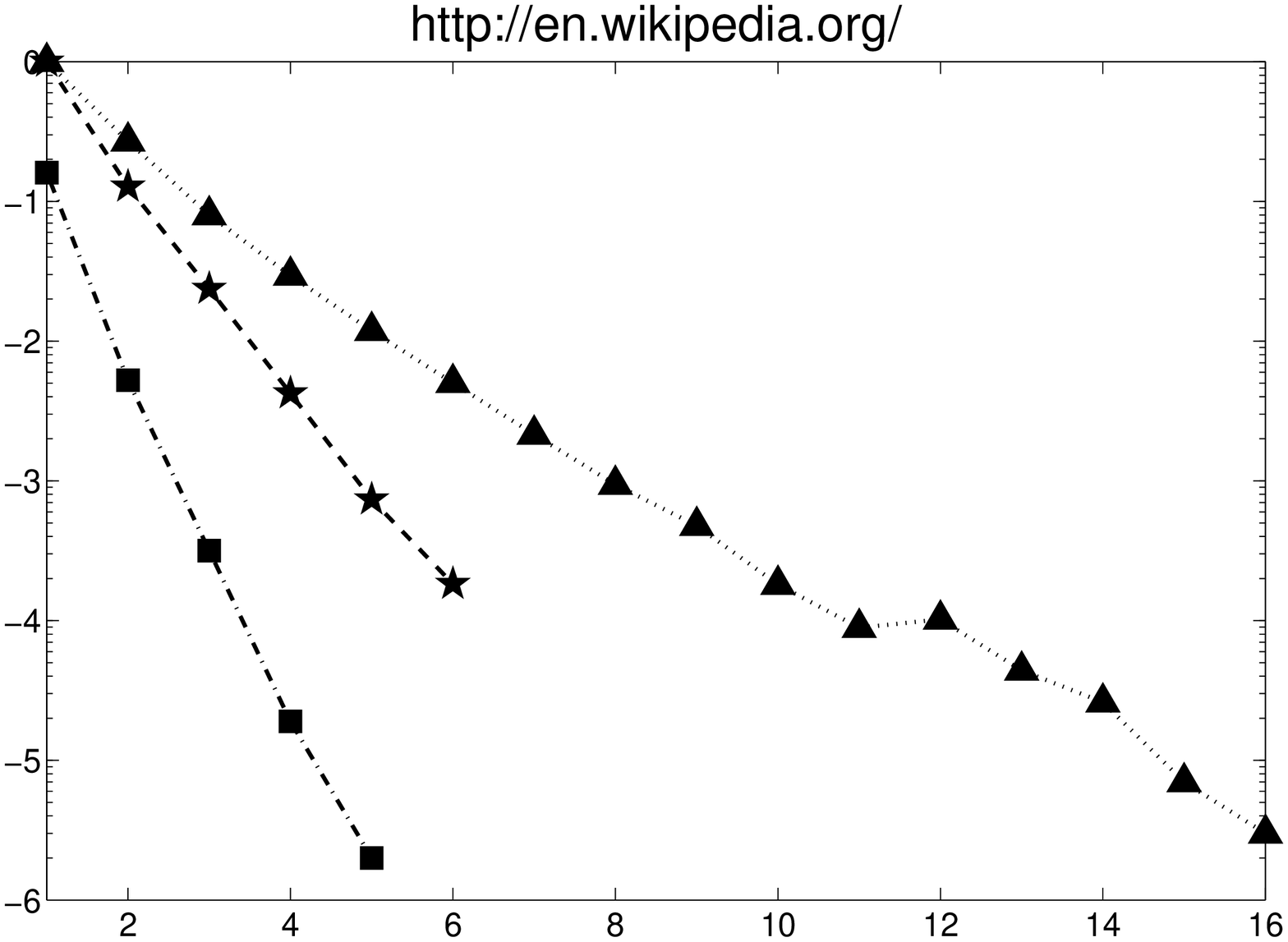}
   \label{wiki}
  }
  \subfigure[]{
   \includegraphics[width=0.3\textwidth]{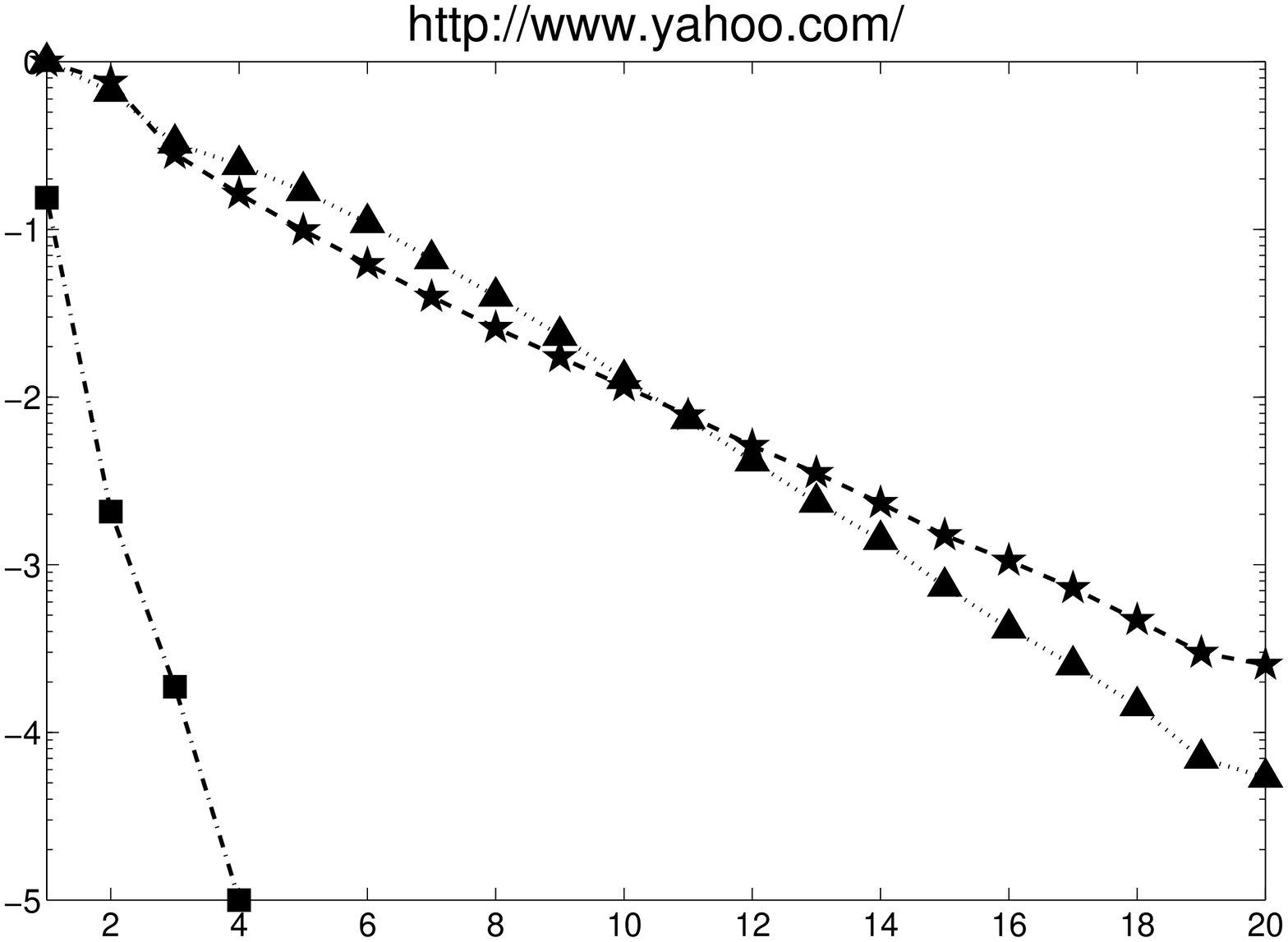}
   \label{yahoo}
  }
  \subfigure[Processing times]{
   \includegraphics[width=0.3\textwidth]{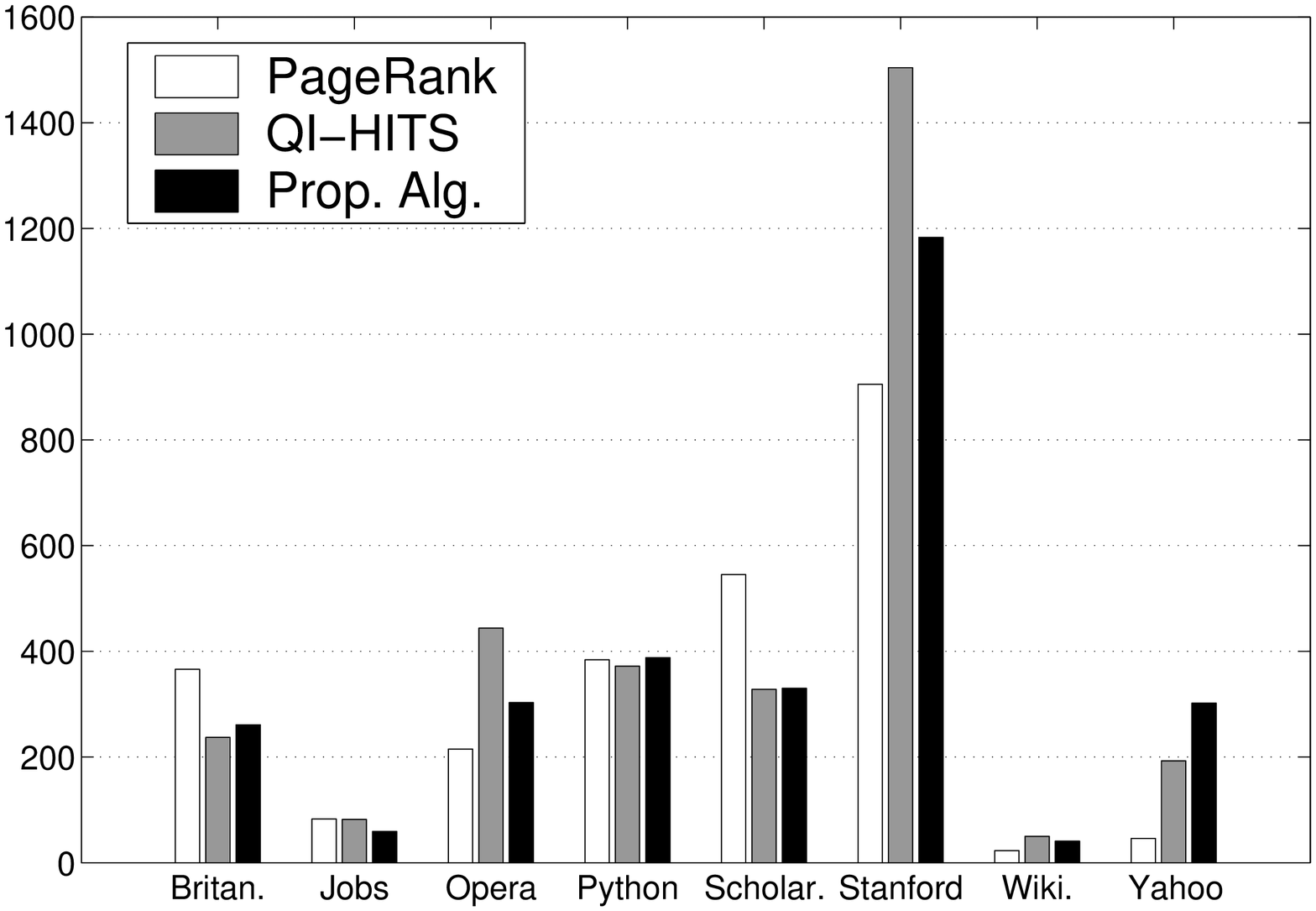}
   \label{barchart}
  }

  \caption{Convergence rates and processing times for original datasets.}
  \label{fig2}
 \end{center}
\end{figure*}

\begin{figure*}
 \begin{center}
 
  \subfigure[]{
   \includegraphics[width=0.3\textwidth]{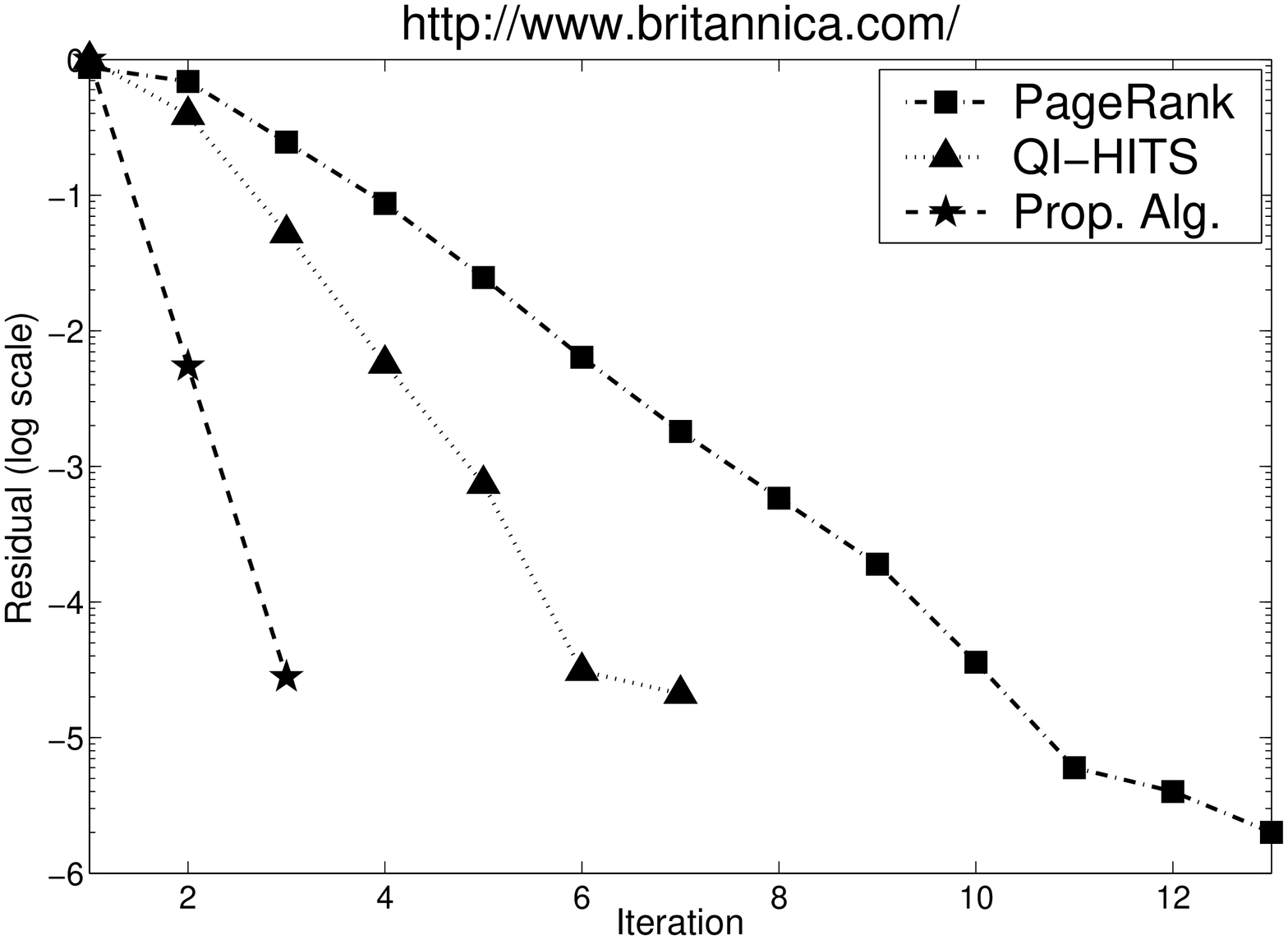}
   \label{britannicabb}
  }
  \subfigure[]{
   \includegraphics[width=0.3\textwidth]{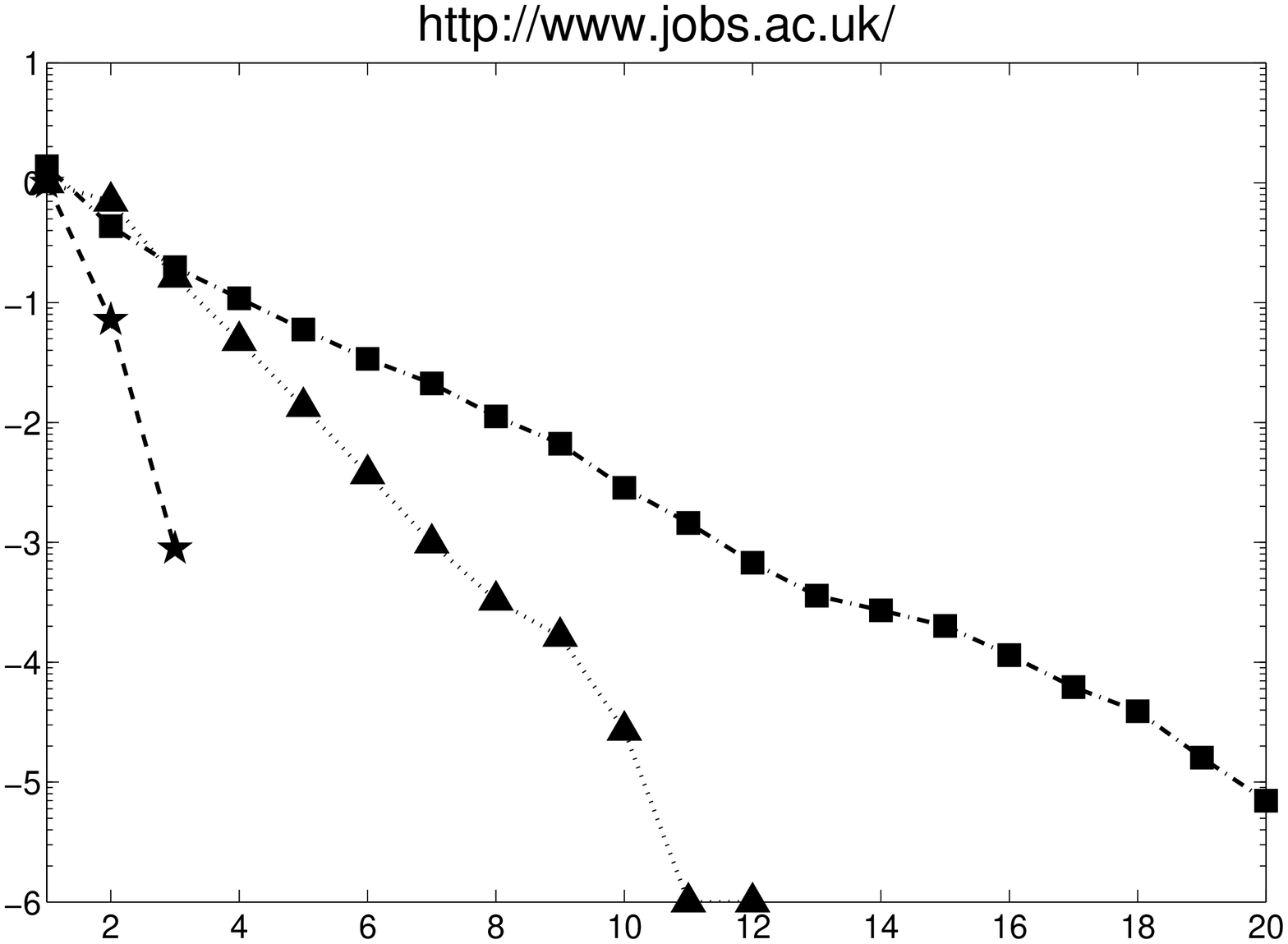}
   \label{jobsbb}
  }
  \subfigure[]{
   \includegraphics[width=0.3\textwidth]{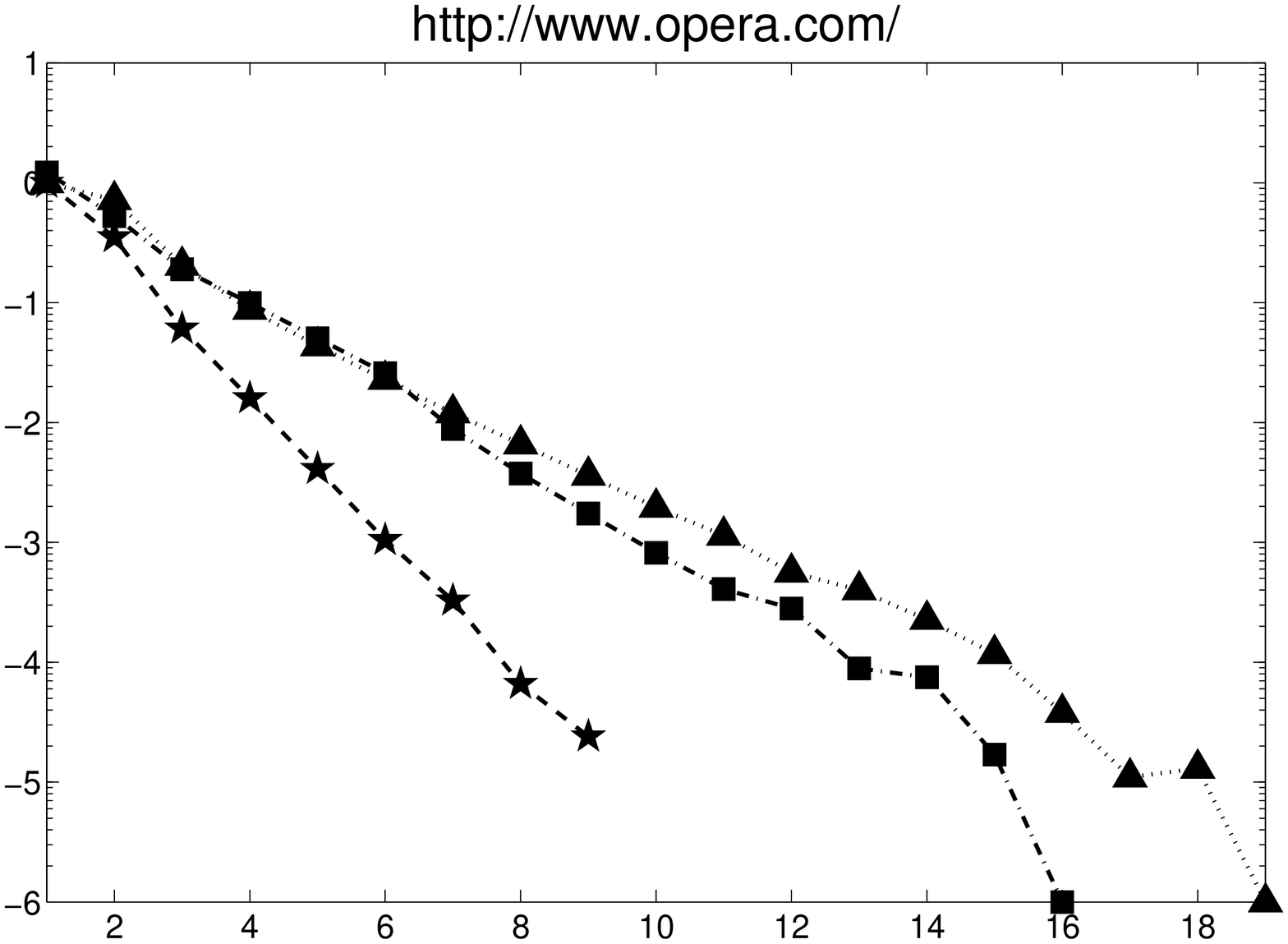}
   \label{operabb}
  }
  \\
 
  \subfigure[]{
   \includegraphics[width=0.3\textwidth]{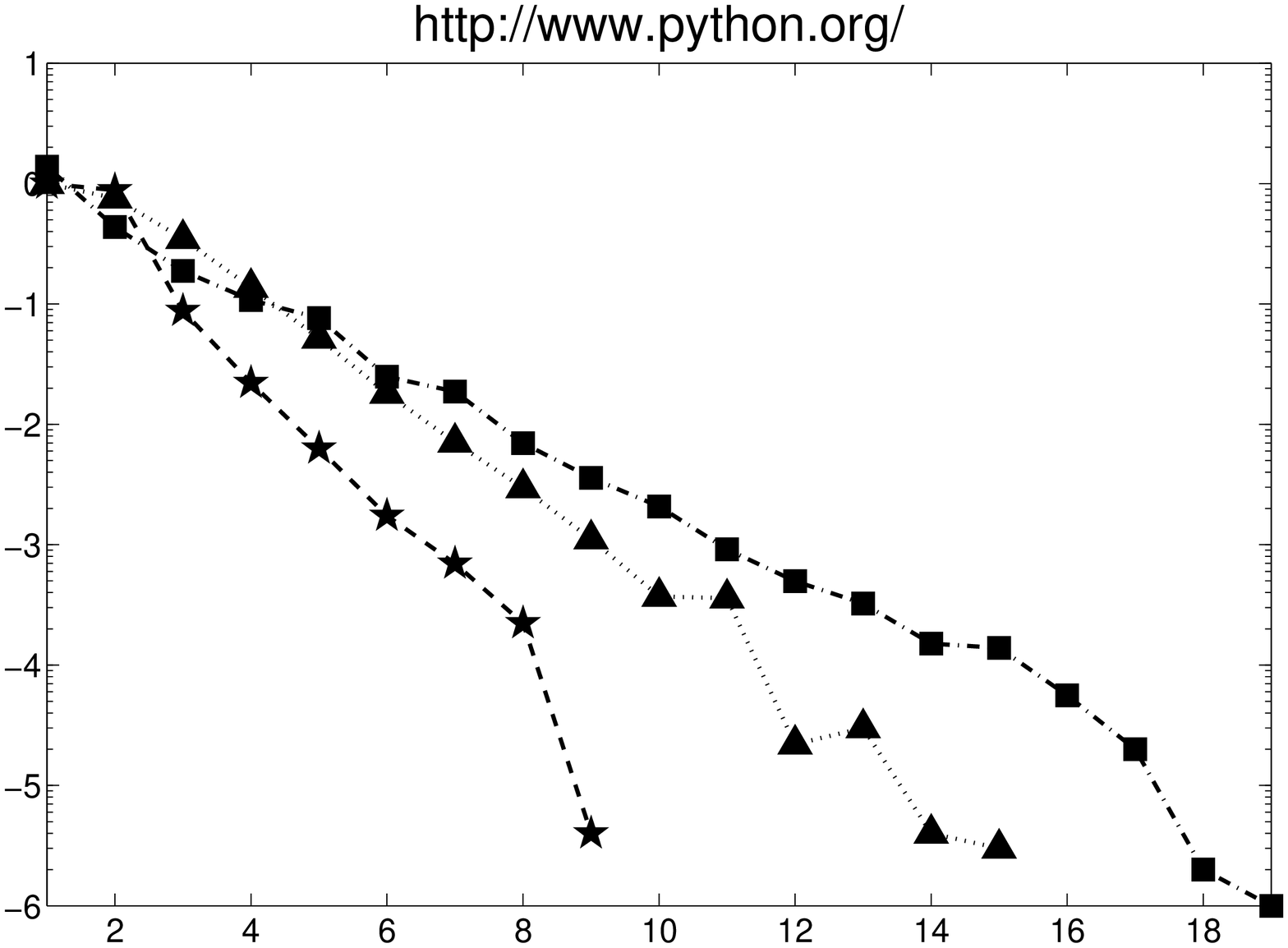}
   \label{pythonbb}
  }
  \subfigure[]{
   \includegraphics[width=0.3\textwidth]{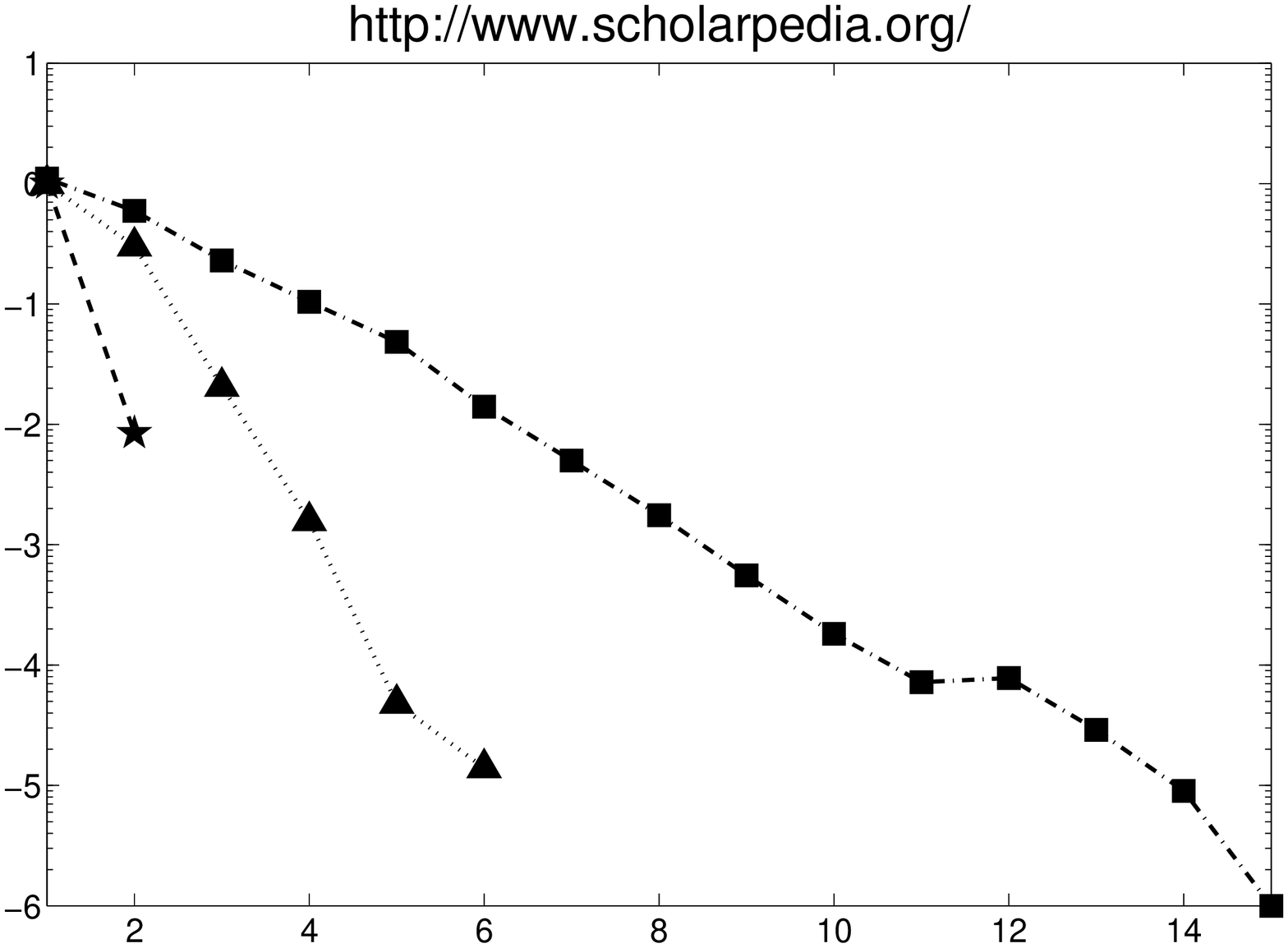}
   \label{scholarpediabb}
  }
  \subfigure[]{
   \includegraphics[width=0.3\textwidth]{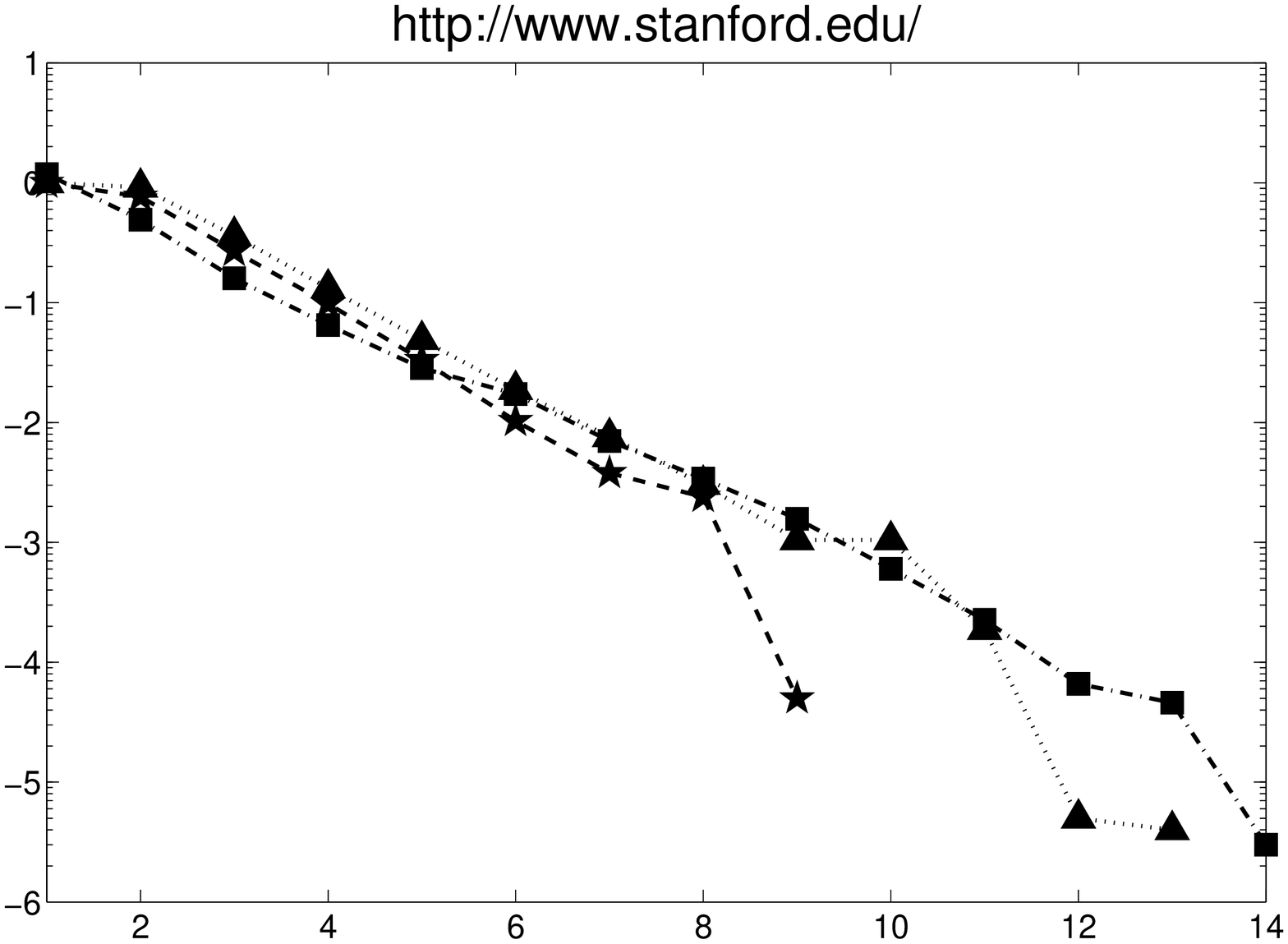}
   \label{stanfordbb}
  }
  \\

  \subfigure[]{
   \includegraphics[width=0.3\textwidth]{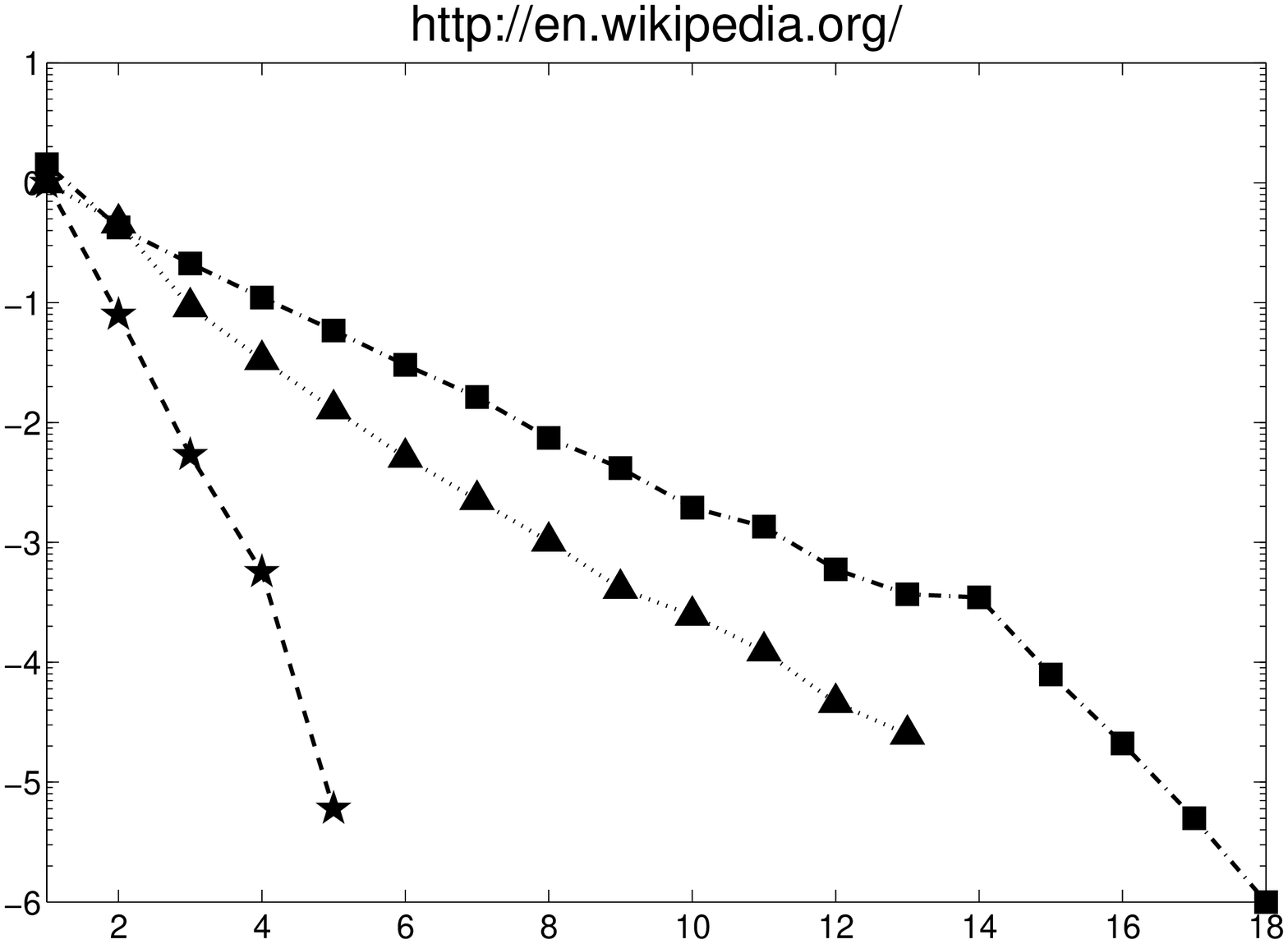}
   \label{wikibb}
  }
  \subfigure[]{
   \includegraphics[width=0.3\textwidth]{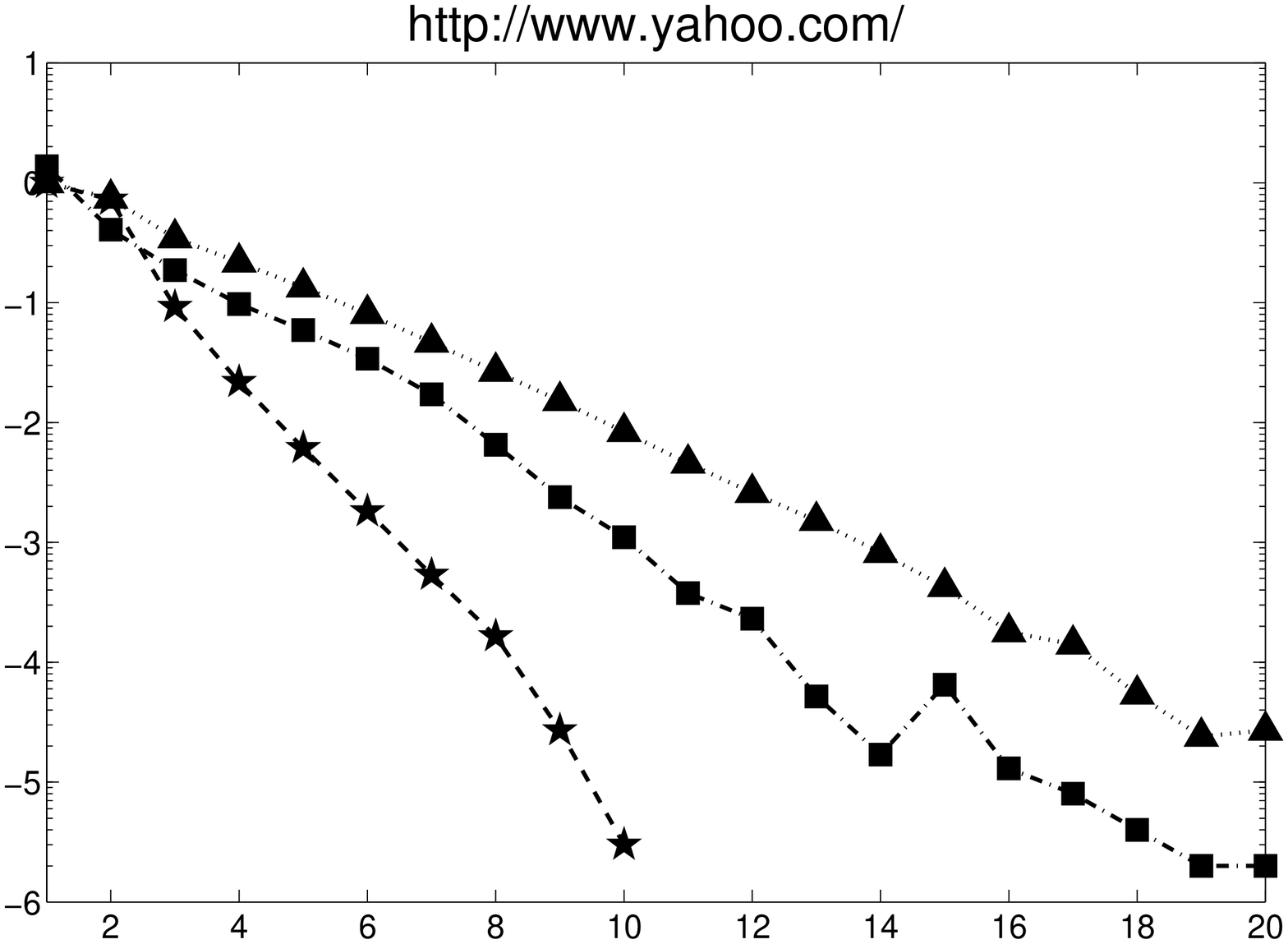}
   \label{yahoobb}
  }
  \subfigure[Processing times]{
   \includegraphics[width=0.3\textwidth]{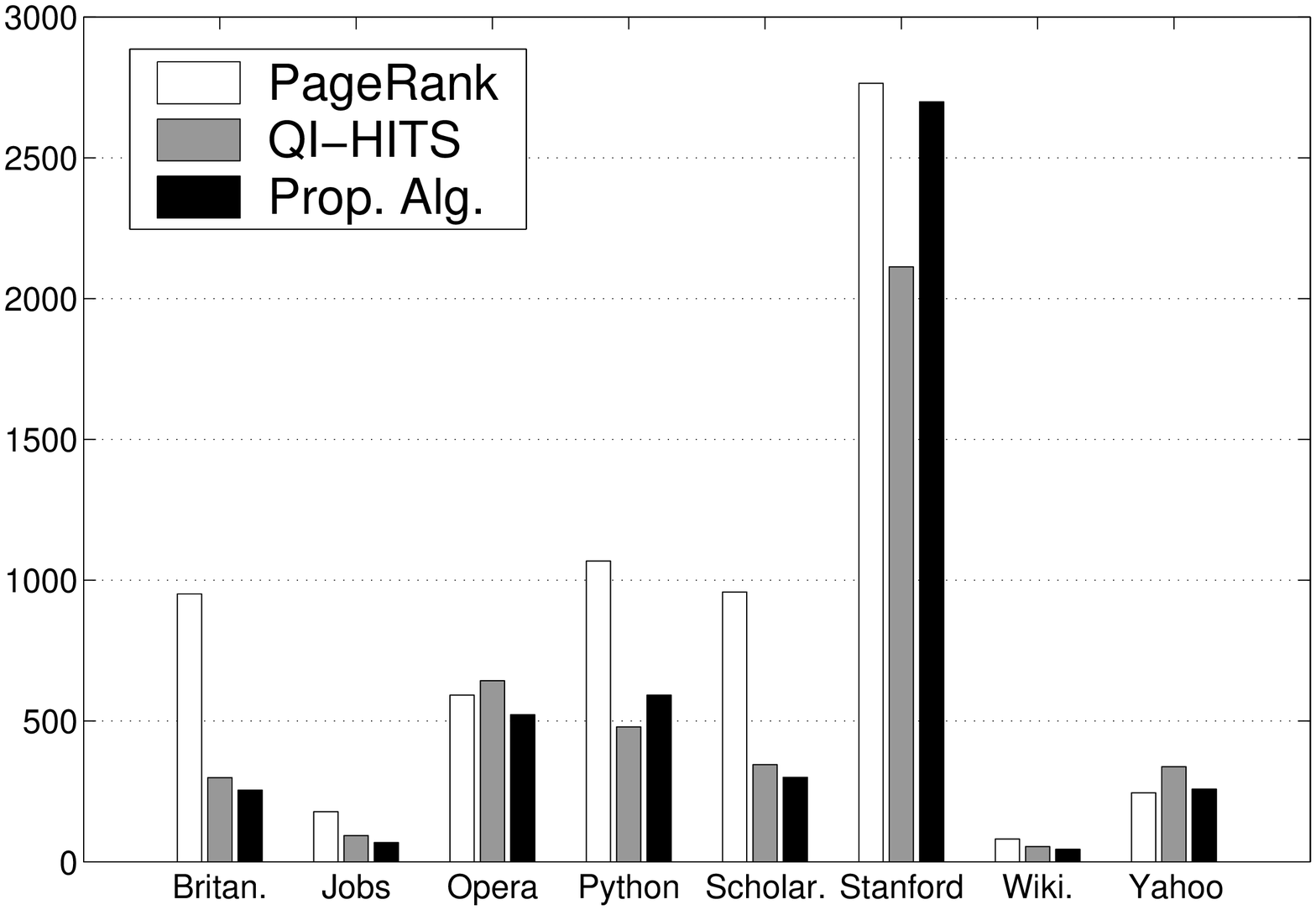}
   \label{barchartbb}
  }

  \caption{Convergence rates and processing times for back button datasets.}
  \label{fig3}
 \end{center}
\end{figure*}

Fig.~\ref{barchart} and \ref{barchartbb} show the processing times (in second) to achieve the same corresponding residual level for the original and the back button datasets respectively. In the original datasets, in average PageRank is the fastest and HITS is the slowest to converge. And in the back button model, the proposed algorithm becomes the fastest in five out of eight cases, PageRank is the slowest in six out of eight cases, and HITS gives moderate performances. The performances of the back button model also agree with the previous works \cite{Kleinberg,Ding,Ding2,Ng,Ng2} where HITS needs less processing time than PageRank.

\subsection{Similarity measures} \label{similarity}
The similarity measures between the authority and hub vectors of QI-HITS and the proposed algorithm are shown in Table \ref{table8}. The purpose of the measurements is to confirm the suitability of the proposed algorithm in approximating the results of QI-HITS. As shown there, the proposed algorithm gives good approximations to the authority vectors, and very good ones to the hub vectors. The measures are the best in the hub vectors of the original datasets. Because Spearman correlation gives about 0.999, it can be conferred that the proposed algorithm returns (almost) exactly the same ordering as QI-HITS. This is not surprising because in the original datasets more than 90\% pages are the dangling pages that have no hub scores.

\subsection{Top pages} \label{toppages}
Table \ref{table9} and \ref{table10} give examples of the results returned by the authority vectors of the three algorithms for wikipedia dataset without query and with query ``programming'' respectively. Note that for brevity only file names are displayed. To get full URLs, each name has to be prefixed with ``http://en.wikipedia.org/wiki/''.
\begin{table*}[t]
 \renewcommand{\arraystretch}{1.3}
  \begin{center}
    \caption{Top 10 results without query for wikipedia dataset.}
    \scriptsize{
    \begin{tabular}{|r|l|l|l|}
     \hline
     No. & PageRank & HITS & Prop.~Alg. \\
     \hline
     1 & \verb|Main_Page| & \verb|Main_Page| & \verb|Main_Page| \\
     2 & \verb|Programming_language| & \verb|Programming_language| & \verb|Programming_language| \\
     3 & \verb|Computer_language| & \verb|C_programming_language| & \verb|2006| \\
     4 & \verb|C_programming_language| & \verb|Java_programming_language| & \verb|C_programming_language| \\
     5 & \verb|Java_programming_language| & \verb|C%2B%2B| & \verb|Operating_system| \\
     6 & \verb|Object-oriented_programming| & \verb|Operating_system| & \verb|Microsoft_Windows| \\
     7 & \verb|Compiler| & \verb|Microsoft_Windows| & \verb|Unix| \\
     8 & \verb|C%2B%2B| & \verb|Object-oriented_programming| & \verb|Linux| \\
     9 & \verb|Operating_system| & \verb|Unix| & \verb|2005| \\
    10 & \verb|Microsoft_Windows| & \verb|Programming_paradigm| & \verb|Java_programming_language| \\    
    \hline
   \end{tabular}}
   \label{table9}
  \end{center}
\end{table*}
\begin{table*}[t]
 \renewcommand{\arraystretch}{1.3}
  \begin{center}
    \caption{Top 10 results with query ``programming'' for wikipedia dataset.}
    \tiny{
    \begin{tabular}{|r|l|l|l|}
     \hline
     No. & PageRank & HITS & Prop.~Alg. \\
     \hline
     1 & \verb|Programming_language| & \verb|Programming_language| & \verb|Programming_language| \\
     2 & \verb|Categorical_list_of_programming_languages| & \verb|Categorical_list_of_programming_languages| & \verb|Categorical_list_of_programming_languages| \\
     3 & \verb|Functional_programming| & \verb|C_programming_language| & \verb|C_programming_language| \\
     4 & \verb|Object-oriented_programming| & \verb|Functional_programming| & \verb|Functional_programming| \\
     5 & \verb|C_programming_language| & \verb|Object-oriented_programming| & \verb|Object-oriented_programming| \\
     6 & \verb|Generic_programming| & \verb|Programming_paradigm| & \verb|Java_programming_language| \\
     7 & \verb|Programming_paradigm| & \verb|Java_programming_language| & \verb|Programming_paradigm| \\
     8 & \verb|Java_programming_language| & \verb|Generic_programming| & \verb|Generic_programming| \\
     9 & \verb|Lisp_programming_language| & \verb|Lisp_programming_language| & \verb|Lisp_programming_language| \\
    10 & \verb|Logic_programming| & \verb|Ada_programming_language| & \verb|Ada_programming_language| \\    
    \hline
   \end{tabular}}
   \label{table10}
  \end{center}
\end{table*}

\section{Conclusions and Future Researches} \label{conclusion}
The proposed algorithm which makes use the definition of authority and hub can be used to accelerate the HITS algorithm. While in original datasets it converges only faster than HITS, in back button datasets it converges faster than both PageRank and HITS. Further, generally there are also some improvements in the processing times, especially for the back button datasets.

The non-uniqueness problem due to the reducibility of the authority matrix $\mathbf{X}$ can be eliminated by forcing the matrix into a positive matrix $\mathbf{\hat{X}}$. This modification not only guarantees the uniqueness, but also tackles the second less obvious problem; producing ranking vectors that inappropriately assign zero scores to some pages.

Based on the similarity measurements, it can be concluded that the vectors produced by the proposed algorithm can be used to approximate the QI-HITS's vectors. And if the QI-HITS vectors are desired instead, the QI-HITS algorithm can be run by using these vectors as the starting vectors for a few last iteration steps. In the case of QI-HITS where the problem involving calculating the stationary vectors of the enormous adjacency matrix of the web graph, even a few number of iteration steps are worth many resources because it takes days to finish the calculations.

There are some interesting future researches related to this work. \emph{First}, as stated in section \ref{introduction} and \ref{relatedworks}, the researches on accelerating the QI-HITS computations are hardly known. The remarkable similarity between the PageRank and QI-HITS formulation \cite{note1} implies that QI-HITS can be accelerated by utilizing the methods discussed in section \ref{relatedworks} in conjunction to the proposed algorithm.

And \emph{second}, as shown in eq.~\ref{eq6}, the HITS algorithm can be accelerated by introducing $\mathbf{Ca}$ and $\mathbf{Ch}$ into the original authority matrix $\mathbf{L}^{T}\mathbf{L}$. While we calculate the entries of $\mathbf{Ca}$ and $\mathbf{Ch}$ based on the HITS definition of authority and hub scores, other schemes like dynamically updating the entries by using the differences between the vectors of current and previous iteration or using previously calculated ranking vectors as the entries probably can also be used \cite{acknowledgement}.

%\newpage
\vspace{25pt}

\begin{profile}
  \Name{Andri Mirzal}
  \Affiliation{Graduate School of Information Science and Technology, Hokkaido University}
  \Address{Kita 14 Nishi 9, Kita-Ku, Sapporo 060-0814, Japan}
  \History{Andri Mirzal is a doctoral course student in Graduate School of Information Science and Technology, Hokkaido University. He received Bachelor of Engineering from Department of Electrical Engineering, Bandung Institute of Technology, and Master of Information Science and Technology from Hokkaido University.}
  \Works{$\bullet$ Complex Network \\
         $\bullet$ Web Search Engine \\
         $\bullet$ Document Clustering}
  %\Membership{$\bullet$ Your Learned Societies}
  %%please place any image files in subdirectory named ``fig''
  \Photo[\epsfig{file=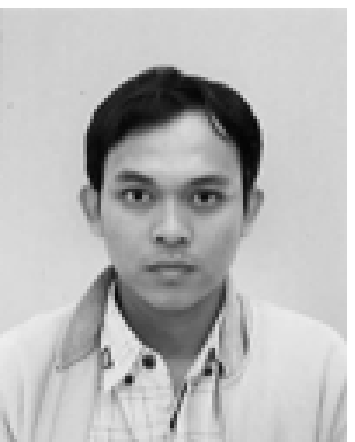,width=70.86pt}]
\end{profile}
\begin{profile}
  \Name{Masashi Furukawa}
  \Affiliation{Graduate School of Information Science and Technology, Hokkaido University}
  \Address{Kita 14 Nishi 9, Kita-Ku, Sapporo 060-0814, Japan}
  \History{Masashi Furukawa received the B.~Sc, M.~Sc, and Ph.~D degrees from Hokkaido University, Japan. He is currently a Professor in the Graduate School of Information Science and Technology, Hokkaido University. During 1970s, he was engaged in the development of 3D CAD/CAM systems. He was a Research Associate participating in the Cornell Injection Molding Project in 1976-1977, and in the Computational Geometry Project of the University of East Anglia in 1981-1982.}
  \Works{$\bullet$ Complex Networks \\
         $\bullet$ Machine Leaning and Evolutionary Computation \\
         $\bullet$ Physics Modeling and Animation}
  \Membership{$\bullet$ Japan Society of Information Processing \\
              $\bullet$ Japan Society of Mechanical Engineers \\
              $\bullet$	Japan Society of Robotics \\
              $\bullet$	Japan Society of Precision Engineering}
  %%please place any image files in subdirectory named ``fig''
  \Photo[\epsfig{file=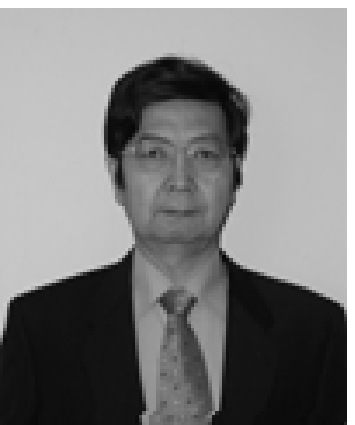,width=70.86pt}]
\end{profile}


\begin{thebibliography}{99}
\bibitem{note1} PageRank and QI-HITS can be written in one vector $\times$ matrix operation, so the tasks of finding the ranking vectors of these algorithms turn to be the problem of calculating dominant eigenvectors of the corresponding matrices. In vector $\times$ matrix operation the PageRank formulation is $\mathbf{p}^{(k+1)T}=\mathbf{p}^{(k)T}\mathbf{G}$ \cite{note3} and the QI-HITS formulation is $\mathbf{a}^{(k+1)T}=\mathbf{a}^{(k)T}\mathbf{L}^{T}\mathbf{L}$. By modifying the authority matrix $\mathbf{L}^{T}\mathbf{L}$ into a positive matrix \cite{note4}, the QI-HITS problem becomes equivalent to the PageRank problem; finding the principal eigenvector of a positive matrix. And after the iteration converges, the hub vector can be revived by calculating $\mathbf{h}^{T} = \mathbf{a}^{T}\mathbf{L}^{T}$
\bibitem{note2} We use ``principal eigenvector'' term if the eigenvalue associated with this vector is the largest, simple and the only eigenvalue on the spectral circle (the matrix is positive), and ``dominant eigenvector'' if the eigenvalue is the largest but may be repeated (the matrix is nonnegative and reducible) or the largest but needs not the only eigenvalue on the spectral circle (the matrix is nonnegative and irreducible).
\bibitem{note3} To ensure a unique and positive PageRank vector exists, the matrix $\mathbf{Do}^{-1}\mathbf{L}$ needs to be modified into a positive matrix \cite{Langville,Farahat}. A proven efficient way of doing this is to adjust the matrix into a stochastic and primitive (irreducibility and aperiodicity imply primitivity) matrix \cite{Langville}. Stochasticity adjustment is done by replacing every zero row (rows associated with the dangling pages) with a probability teleportation vector. And primitivity adjustment is done by replacing every zero entry with a small positive real number. The stochastic and primitive version of $\mathbf{Do}^{-1}\mathbf{L}$ is the positive Google matrix $\mathbf{G}=\alpha\big(\mathbf{Do}^{-1}\mathbf{L}+\mathbf{d}(1/N\,\mathbf{e}^{T})\big)+(1-\alpha)1/N\,\mathbf{e}\,\mathbf{e}^{T}$. Further, the power method applied to $\mathbf{p}^{(k+1)T} = \mathbf{p}^{(k)T}\mathbf{G}$ is guaranteed to converge to the PageRank vector $\mathbf{p}$ as long as the starting vector is not in the range $(\mathbf{G}^{T}-\lambda_{1}\mathbf{I})$ \cite{Langville}.
\bibitem{note4} To ensure the uniqueness of the authority and hub vectors, the authority and hub matrices must be modified into positive matrices \cite{Langville,Farahat}. Let $\mathbf{\hat{A}}$ and $\mathbf{\hat{H}}$ be the positive version of the authority matrix and hub matrix respectively. We can define them as $\mathbf{\hat{A}} = \zeta\,\mathbf{L}^{T}\mathbf{L}+(1/N)(1-\zeta)\,\mathbf{e}\,\mathbf{e}^T$, and $\mathbf{\hat{H}} = \zeta\,\mathbf{L}\,\mathbf{L}^{T}+(1/N)(1-\zeta)\,\mathbf{e}\,\mathbf{e}^T$, where $0<\zeta<1$ is a constant that should be set near to $1$ to preserve the hyperlink structure information.
\end{thebibliography}
\end{document}